\documentclass[amsmath,amssymb,prb,showpacs,twocolumn,superscriptaddress]{revtex4}

\usepackage{graphicx}
\usepackage{color}

\begin{document}

\title{Inelastic effects in molecular junctions
       in the Coulomb and Kondo regimes:\\ 
       Nonequilibrium equation-of-motion approach.}
\date{\today}
\author{Michael Galperin}
\affiliation{Department of Chemistry and Nanotechnology Center,
   Northwestern University, Evanston IL 60208}
\author{Abraham Nitzan}
\affiliation{School of Chemistry, The Sackler Faculty of Science,
   Tel Aviv University, Tel Aviv 69978, Israel}
\author{Mark A. Ratner}
\affiliation{Department of Chemistry and Nanotechnology Center,
   Northwestern University, Evanston IL 60208}

\begin{abstract}
Inelastic effects in the Coulomb blockade and Kondo regimes of electron 
transport
through molecular junctions are considered within a simple nonequilibrium
equation-of-motion (EOM) approach. The scheme is self-consistent, and 
can qualitatively reproduce the main experimental observations
of vibrational features in Coulomb blockade 
[H.~Park~et~al., Nature \textbf{407}, 57 (2000)] 
and Kondo [L.~H.~Yu~et~al., Phys. Rev. Lett. \textbf{93}, 266802 (2004)] 
regimes.
Considerations similar to the equilibrium EOM approach by Meir~et~al. 
[Phys.~Rev.~Lett. \textbf{66}, 3048 (1991); ibid. \textbf{70}, 2601 (1993)]
are used on the Keldysh contour to account for the nonequilibrium nature of the 
junction, and dressing by appropriate Franck-Condon (FC) factors 
is used to account for vibrational features.
Results of the equilibrium EOM scheme by 
Meir~et~al.
are reproduced in the appropriate limit.
\end{abstract}

\maketitle

\section{\label{intro}Introduction}
Fast development of experimental techniques in the area of molecular 
electronics makes it possible to observe the response of molecular conduction 
junctions in a wide range of external parameters, 
such as source-drain and gate voltages.\cite{Reed}
Coulomb blockade (that characterizes the weak molecule-lead coupling
limit), where transport
through the molecular junction is suppressed due to high
charging energy, and Kondo effect (encountered at sufficiently low
temperature and strong molecule-lead coupling),
when a correlation between localized (molecular) and band (contacts) electrons 
manifests itself in molecular junctions as a maximum in electrical conductance 
near $V_{sd}\sim 0$, were observed in the $I/V_{sd}$
characteristics of such junctions.\cite{McEuen,Park,Bao,Bjornholm,Natelson,Zant}
These are often accompanied by vibrational features that result from
coupling between electronic and vibrational degrees of freedom.
The latter can be associated with molecular center-of-mass motion\cite{Park_C60}
or with intra molecular vibrations.\cite{McEuen,Bao,Natelson}

Early theoretical approaches to transport in the 
Coulomb blockade regime were based either on linear response theory
for near equilibrium situations\cite{Beenakker,MeirWingreenLee_CB,Kang,Ness} 
or by treating transport at the level of quasi classical rate 
equations.\cite{Gurvitz,GhoshDatta_rate}
While the second approach to nonequilibrium transport is justified in the case 
of pure Coulomb blockade (where hopping between molecule and contacts 
is rare), the intermediate regime, e.g. the case of stronger molecule-leads
coupling relevant for observation of nonequilibrium Kondo resonance, should be
treated at a more sophisticated level. Recent approaches dealing with
nonequilibrium Coulomb blockade and/or Kondo effect are based either on the
slave-boson technique\cite{LangrethNordlander,WingreenMeir,Krawiec,Guo}, 
the equation-of-motion method\cite{Krawiec,Ng,NiuLinLin,Swirkowicz}, 
the Fock-space rate equation scheme\cite{GhoshDatta_rate},
or the contour perturbation 
theory.\cite{Flores,Paaske,Glazman,Kang,Ueda,KomnikGogolin,Schoeller,Hamasaki}
Inelastic effects were not considered in the references above.

Here we present a simple generalization of the equilibrium equation-of-motion 
approach used in the Coulomb\cite{MeirWingreenLee_CB,HaugJauho} regime 
(applied later also to the Kondo\cite{MeirWingreenLee_Kondo} situation) 
to the case of nonequilibrium transport. 
The main difference between our approach and earlier nonequilibrium EOM
studies\cite{Krawiec,NiuLinLin,Swirkowicz}
is a simple appealing structure of the Green function,
the evaluation of which (in the absence of electron-phonon coupling)
does not require a time-consuming self-consistent procedure.
As was indicated earlier,\cite{HaugJauho} this Green function
expression reduces to the exact solution both for an isolated 
molecule and in the limit of noninteracting electrons. 
We also generalize this basic scheme to include inelastic effects 
approximately, within an approach based on the Born-Oppenheimer 
approximation that is commonly used in Marcus theory of electron 
transfer.\cite{Marcus}
Numerical calculations are performed and qualitative correspondence to 
experimental data is demonstrated.

Our model and theoretical procedure are presented in Section~\ref{model}.
Numerical results for the Coulomb blockade regime are given and discussed in 
Section~\ref{results}. The Kondo regime is discussed in Section~\ref{kondo}.
Section~\ref{conclude} concludes.

\section{\label{model}Model and method}
We describe the molecular junction within a single resonant level (molecular 
electronic orbital) model, with electron-electron on-site repulsion 
(Hubbard term) and polaronic coupling to a local vibrational mode. 
The latter is coupled to a bosonic thermal bath. 
The electronic orbital is coupled to two ($L$ and $R$) free-electron
reservoirs representing the leads, each at its own equilibrium.

The corresponding Hamiltonian is
\begin{align}
\label{H}
 \hat H &= \sum_{K=L,R}\sum_{k\in K,\sigma}
   \varepsilon_{k\sigma}\hat c^\dagger_{k\sigma}\hat c_{k\sigma}
 + \sum_\sigma\varepsilon_\sigma\hat d^\dagger_\sigma\hat d_\sigma 
 + \omega_0\hat a^\dagger\hat a
 \nonumber \\
 &+ \sum_\beta \omega_\beta \hat b^\dagger_\beta\hat b_\beta
 + \sum_{K=L,R}\sum_{k\in K,\sigma}
    \left(V_{k\sigma}\hat c^\dagger_{k\sigma}\hat d_\sigma + \mbox{H.c.}\right) 
 \\
  &+ U \hat n_{\uparrow}\hat n_{\downarrow}
  + M \hat Q_a \sum_\sigma \hat n_\sigma 
  + \sum_\beta U_\beta \hat Q_a\hat Q_\beta
 \nonumber
\end{align}
where $\sigma=\uparrow,\downarrow$ is the electron spin index,
$\hat c_{k\sigma}$ ($\hat c^\dagger_{k\sigma}$) are destruction (creation)
operators for electronic state $k\sigma$ in the contacts,
$\hat d_\sigma$ ($\hat d^\dagger_\sigma$) destroys (creates) electron in the
molecular orbital, $\hat a$ ($\hat a^\dagger$) are second quantization 
operators for the
local vibrational mode, and $\hat b_\beta$ ($\hat b^\dagger_\beta$) are
the corresponding boson operators for thermal bath modes.
Also
\begin{equation}
 \hat Q_a = \hat a + \hat a^\dagger \qquad
 \hat Q_\beta = \hat b_\beta + \hat b_\beta^\dagger
\end{equation}
are displacement operators for corresponding modes and 
$\hat n_\sigma=\hat d^\dagger_\sigma \hat d_\sigma$ .
Here and below we use $\hbar=1$ and $e=1$.
After small polaron (canonical or Lang-Firsov) transformation\cite{Mahan}
the Hamiltonian takes the form (for details see Ref.~\onlinecite{strong_elph})
\begin{align}
\label{barH}
 \hat{\bar H} &= \sum_{K=L,R}\sum_{k\in K,\sigma}
   \varepsilon_{k\sigma}\hat c^\dagger_{k\sigma}\hat c_{k\sigma}
 + \sum_\sigma\bar\varepsilon_\sigma\hat d^\dagger_\sigma\hat d_\sigma
 + \omega_0\hat a^\dagger\hat a
 \nonumber \\
 &+ \sum_\beta \omega_\beta \hat b^\dagger_\beta\hat b_\beta
 + \sum_{K=L,R}\sum_{k\in K,\sigma}
    \left(\bar V_{k\sigma}\hat c^\dagger_{k\sigma}\hat d_\sigma 
         + \mbox{H.c.}\right)
 \\
  &+ \bar U \hat n_{\uparrow}\hat n_{\downarrow}
  + \sum_\beta U_\beta \hat Q_a\hat Q_\beta
 \nonumber
\end{align}
where
\begin{align}
 \bar\varepsilon_\sigma &= \varepsilon_\sigma - M^2/\omega_0
 \\
 \bar U &= U - 2 M^2/\omega_0
 \\
 \label{Vks}
 \bar V_{k\sigma} &= V_{k\sigma} \hat X_a
\end{align}
and where
\begin{equation}
 \label{Xa}
 \hat X_a = \exp\left(i\lambda_a\hat P_a\right);
 \qquad \lambda_a=\frac{M}{\omega_0}
\end{equation}
is the phonon shift generator operator with
\begin{equation}
 \label{Pa}
 \hat P_a = -i\left(\hat a - \hat a^\dagger\right)
\end{equation}
$\hat P_a$, Eq.(\ref{Pa}), is the phonon momentum operator;
we use the term phonon to characterize both molecular and bath vibrations.

The Hamiltonian (\ref{barH}) is our starting point for the calculation of the
steady-state current across the junction, using the
nonequilibrium Green function (NEGF)
expression derived in Refs.~\onlinecite{HaugJauho,current}
\begin{equation}
 \label{current}
 I_K = \frac{e}{\hbar} \sum_\sigma \int\frac{dE}{2\pi}
 \left[ \Sigma_{K,\sigma}^{<}(E)\, G_\sigma^{>}(E) 
      - \Sigma_{K,\sigma}^{>}(E)\, G_\sigma^{<}(E) \right]
\end{equation}
Here $\Sigma_{K,\sigma}^{<,>}$ are lesser/greater projections of the 
self-energy due to coupling to the contact $K$ ($K=L,R$)
\begin{eqnarray}
 \label{SEKlt}
 \Sigma_{K,\sigma}^{<}(E) &=& i f_K(E) \Gamma_{K,\sigma}(E) \\
 \label{SEKgt}
 \Sigma_{K,\sigma}^{>}(E) &=& -i [1-f_K(E)] \Gamma_{K,\sigma}(E)
\end{eqnarray}
with $f_K(E)$ the Fermi distribution in the contact $K$ and
\begin{equation}
 \label{GammaK}
 \Gamma_{K,\sigma}(E) = 2\pi \sum_{k\in K} |V_{k\sigma}|^2 
 \delta(E-\varepsilon_k)
\end{equation}

The lesser and greater Green functions in (\ref{current}) are
Fourier transforms to energy space of projections onto
the real time axis of the electron Green function on the Keldysh contour
\begin{align}
 \label{GFKeldysh}
 G_\sigma(\tau_1,\tau_2)
 &= -i<T_c \hat d_\sigma(\tau_1)\hat d_\sigma^\dagger(\tau_2)>_H
 \\
 &= -i<T_c \hat d_\sigma(\tau_1)\hat X_a(\tau_1)\,
          \hat d_\sigma^\dagger(\tau_2)\hat X_a^\dagger(\tau_2)>_{\bar H}
 \nonumber
\end{align}
where the subscripts $H$ and $\bar H$ indicate which Hamiltonian,
(\ref{H}) or (\ref{barH}) respectively,
determines evolution of the system, and $T_c$ is the contour ordering
operator. In what follows we use the second form and will drop the
subscript $\bar H$ while keeping in mind that time evolution is determined
by the Hamiltonian (\ref{barH}). We next decouple electron and phonon dynamics
in the spirit of the Born-Oppenheimer theory within the Condon approximation
\begin{equation}
 \label{appGFKeldysh}
 G_\sigma(\tau_1,\tau_2) \approx G_\sigma^{(e)}(\tau_1,\tau_2)\, K(\tau_1,\tau_2)
\end{equation}
where
\begin{eqnarray}
 \label{defGe}
 G_\sigma^{(e)}(\tau_1,\tau_2) &=& 
 -i<T_c \hat d_\sigma(\tau_1)\hat d_\sigma^\dagger(\tau_2)>
 \\
 \label{K}
 K(\tau_1,\tau_2) &=& <T_c \hat X_a(\tau_1)\hat X_a^\dagger(\tau_2)>
\end{eqnarray}

The shift generator correlation function $K$ can be expressed within 
the second order cumulant expansion in terms of the phonon Green function
(for derivation see Ref.~\onlinecite{strong_elph})
\begin{align}
 \label{XXKeldysh}
 &K(\tau_1,\tau_2) =
 \exp\left\{\lambda_a^2\left[i D_{P_aP_a}(\tau_1,\tau_2)
                -<\hat P_a^2>\right]\right\}
 \\
 &D_{P_aP_a}(\tau_1,\tau_2) = -i<T_c \hat P_a(\tau_1)\hat P_a(\tau_2)>
\end{align}
while the phonon Green function $D$ obeys approximately an equation which 
resembles the usual Dyson equation
\begin{align}
 \label{DKeldysh}
 &D_{P_aP_a}(\tau,\tau') = D_{P_aP_a}^{(0)}(\tau,\tau')
 \\
 &+ \int_c d\tau_1 \int_c d\tau_2\, D_{P_aP_a}^{(0)}(\tau,\tau_1)\,
 \Pi_{P_aP_a}(\tau_1,\tau_2)\, D_{P_aP_a}(\tau_2,\tau')
 \nonumber
\end{align}
with 
\begin{align}
 \label{DSEKeldysh}
 &\Pi_{P_aP_a}(\tau_1,\tau_2) = \sum_\beta |U_\beta|^2
 D_{P_\beta P_\beta}(\tau_1,\tau_2)
 \\ 
 &- i\lambda_a^2\sum_{k\in\{L,R\},\sigma}|V_{k\sigma}|^2\left[
     g_{k,\sigma}(\tau_2,\tau_1)G_\sigma^{(e)}(\tau_1,\tau_2)
     K(\tau_1,\tau_2)\right.
 \nonumber\\
  &\left.\quad\qquad\qquad\qquad\qquad +  (\tau_1\leftrightarrow\tau_2)\right]
 \nonumber
\end{align}
the analog of a self-energy. $g_{k,\sigma}$ is the free electron GF in the 
contact, defined in (\ref{defg}) below.

To obtain an expression for the Green function $G_\sigma^{(e)}$ 
we follow the equation-of-motion (EOM) method of
Meir, Wingreen, and Lee\cite{MeirWingreenLee_CB,HaugJauho} where it
was applied for a near-equilibrium situation,
except that we consider the EOMs on the Keldysh contour 
in order to take into account the nonequilibrium condition.
In the spirit of the Born-Oppenheimer approximation we regard the shift
generator operators $\hat X_a$ as parameters incorporated into transfer
matrix elements $\bar V_{k\sigma}$. 
The solution of the electronic problem is thus carried out as in the
absence of electron-phonon coupling\cite{MeirWingreenLee_CB,HaugJauho}
with renormalized parameters $U$ (${}\to\bar U$) and $V$ (${}\to\bar V$).
The result is then averaged over the phonon subspace. This average is
obviously not needed in the absence of electron-phonon coupling, $M=0$,
in which case $G_\sigma=G_\sigma^{(e)}$. 
This leads to (for derivation see Appendix~\ref{A})
\begin{align}
 \label{Ge}
 G_\sigma^{(e)}(\tau_1,\tau_2) &= 
 \left[1-<\hat n_{\bar\sigma}>\right] G_{2,\sigma}^{(e)}(\tau_1,\tau_2)
 \\ &+
 <\hat n_{\bar\sigma}> G_{3,\sigma}^{(e)}(\tau_1,\tau_2)
 \nonumber
\end{align}
where the GFs $G_{i,\sigma}^{(e)}$ ($i=\{1,2,3,4\}$) obey
\begin{equation}
 \label{Gei}
 \int_c d\tau\, 
 \hat G^{-1}_{i,\sigma}(\tau_1,\tau) G_{i,\sigma}^{(e)}(\tau,\tau_2)
 = \delta(\tau_1,\tau_2)
\end{equation}
with
\begin{align}
 \label{Ge1_m1}
 \hat G^{-1}_{1,\sigma}(\tau,\tau') =&
 \left[\delta(\tau,\tau')\left(i\frac{\partial}{\partial\tau}
                               -\varepsilon_\sigma-U\right)
 \right.\\&\left.
      -\Sigma_{\sigma 0}(\tau,\tau')-\Sigma_{\sigma 3}(\tau,\tau') \right]
 \nonumber\\
 \label{Ge2_m1}
 \hat G^{-1}_{2,\sigma}(\tau,\tau') =& 
 \left[\delta(\tau,\tau')\left(i\frac{\partial}{\partial\tau}-\varepsilon_\sigma\right)
      -\Sigma_{\sigma 0}(\tau,\tau')
 \right. \\ &\quad +\left. U\int_c d\tau''\,
      G_{1,\sigma}^{(e)}(\tau,\tau'')\Sigma_{\sigma 1}(\tau'',\tau') \right]
 \nonumber \\
 \label{Ge3_m1}
 \hat G^{-1}_{3,\sigma}(\tau,\tau') =&
 \left[\delta(\tau,\tau')\left(i\frac{\partial}{\partial\tau}-\varepsilon_\sigma-U\right)
      -\Sigma_{\sigma 0}(\tau,\tau')
 \right.\\ &\quad -\left. U\int_c d\tau''\,
      G_{4,\sigma}^{(e)}(\tau,\tau'')\Sigma_{\sigma 2}(\tau'',\tau') \right]
 \nonumber \\
 \label{Ge4_m1}
 \hat G^{-1}_{4,\sigma}(\tau,\tau') =&
 \left[\delta(\tau,\tau')\left(i\frac{\partial}{\partial\tau}-\varepsilon_\sigma\right)
 \right. \\&\left.
      -\Sigma_{\sigma 0}(\tau,\tau')-\Sigma_{\sigma 3}(\tau,\tau') \right]
 \nonumber
\end{align}
Expressions for `self-energies' $\Sigma_{\sigma i}$ ($i=\{0,1,2,3\}$)
are given by
\begin{align}
 \label{SEs0}
 \Sigma_{\sigma 0}(\tau,\tau') &= \sum_k \left|V_{k\sigma}\right|^2
    g_{k,\sigma}(\tau,\tau') <T_c\hat X_a^\dagger(\tau)\,\hat X_a(\tau')>
 \\
 \label{SEs1}
 \Sigma_{\sigma 1}(\tau,\tau') &= \sum_k <\hat n_{k\bar\sigma}>
 \nonumber \\&\times
 \left[
 \left|V_{k\bar\sigma}\right|^2 g_{k,\bar\sigma}^{(1)}(\tau,\tau')
 <T_c \hat X_a(\tau)\,\hat X_a^\dagger(\tau')>
 \right. \\ & \left. +
 |V_{k\bar\sigma}|^2 g_{k,\bar\sigma}^{(2)}(\tau,\tau')
 <T_c\hat X_a^\dagger(\tau)\,\hat X_a(\tau')>
 \right]
 \nonumber \\
 \label{SEs2}
 \Sigma_{\sigma 2}(\tau,\tau') &= \Sigma_{\sigma 3}(\tau,\tau')
 - \Sigma_{\sigma 1}(\tau,\tau')
 \\
 \label{SEs3}
 \Sigma_{\sigma 3}(\tau,\tau') &= \sum_k
  \left[
 \left|V_{k\bar\sigma}\right|^2 g_{k,\bar\sigma}^{(1)}(\tau,\tau')
 <T_c \hat X_a(\tau)\,\hat X_a^\dagger(\tau')>
 \right. \nonumber \\ & \left. +
 \left|V_{k\bar\sigma}\right|^2 g_{k,\bar\sigma}^{(2)}(\tau,\tau')
 <T_c\hat X_a^\dagger(\tau)\,\hat X_a(\tau')>
 \right]
\end{align}
with $\bar\sigma$ denoting the spin opposite to $\sigma$.
The free electron propagators $g_{k,\sigma}$ and $g_{k,\bar\sigma}^{(j)}$,
$j=1,2$ are defined by
\begin{align}
 \label{defg}
 &\left[i\frac{\partial}{\partial\tau}-\varepsilon_{k\sigma}\right]
 g_{k,\sigma}(\tau,\tau') = \delta(\tau,\tau')
 \\
 \label{defg1}
 &\left[i\frac{\partial}{\partial\tau}+\varepsilon_{k\bar\sigma}
  -\varepsilon_\sigma-\varepsilon_{\bar\sigma}-U\right]
 g_{k,\bar\sigma}^{(1)}(\tau,\tau') = \delta(\tau,\tau')
 \\
 \label{defg2}
 &\left[i\frac{\partial}{\partial\tau}-\varepsilon_{k\bar\sigma}
 -\varepsilon_\sigma+\varepsilon_{\bar\sigma}\right]
 g_{k,\bar\sigma}^{(2)}(\tau,\tau') = \delta(\tau,\tau')
\end{align}
For $M=0$, $\bar V$ Franck-Condon (FC) factors 
(i.e. shift generator correlation
functions $<X\, X^\dagger>$ and $<X^\dagger\, X>$) should be taken as 1
in (\ref{SEs0})-(\ref{SEs3}). Below the SEs in this case will be denoted
$\Sigma_{\sigma j}^{(e)}$ ($j=0,1,2,3$).
Note that the retarded projections of these are equivalent to
the SEs introduced in Ref.~\onlinecite{MeirWingreenLee_CB}. 
For example, taking the retarded projection of (\ref{SEs3}) and
Fourier transforming to energy space leads to
\begin{align}
 \Sigma_{\sigma 3}^{(e)r}(E) = \sum_{k\in L,R} \left|V_{k\bar\sigma}\right|^2
 &\left[
  \frac{1}
   {E+\varepsilon_{k\bar\sigma}-\varepsilon_\sigma-\varepsilon_{\bar\sigma}-U}
 \right.\nonumber \\&\left.
 +\frac{1}
  {E-\varepsilon_{k\bar\sigma}-\varepsilon_\sigma+\varepsilon_{\bar\sigma}}
 \right]
\end{align}
which is identical to Eq.(9) in Ref.~\onlinecite{MeirWingreenLee_CB} for $i=3$.
Other expressions are obtained in a similar way.

Consider first the case with no electron-phonon coupling.
The structure of expression (\ref{Ge}) for the nonequilibrium GF 
$G_\sigma^{(e)}$ is appealingly simple and has two important implications.
First, it provides a convenient way for handling the Hubbard repulsion term $U$.
While the case of weak electron-electron interaction can be handled
by taking this term as a perturbation,\cite{Ueda} the case of strong
interaction cannot be handled in this way, but including $U$ in $H_0$
makes standard diagrammatic techniques 
unusable.\cite{X_footnote}
This difficulty is circumvented by Eq.(\ref{Ge}), that expresses the system GF
as a superposition (with the level population $n$ defining weight parameters)
of simpler GFs associated with Hamiltonians that
do not depend on $U$ (apart from a parametric energy shift) for which the 
Wick's theorem is applicable.
Secondly, by using the EOM method on the Keldysh contour we are able
to derive not only the retarded GF as in Ref.~\onlinecite{MeirWingreenLee_CB}
but also the other projections, in particular the lesser GF that can be
used to evaluate the level populations
\begin{equation}
 \label{defnsigma}
 <\hat n_{\sigma}>=-i/2\pi \int dE\, G^{(e)<}_{\sigma}(E)
\end{equation}
This, together with Eq.(\ref{Ge}), lead to an explicit expression for 
$<\hat n_{\sigma}>$. Denoting
\begin{equation}
 I_{i,\sigma}=-i/2\pi\int dE\, G^{(e)<}_{i,\sigma}(E)
\end{equation}
one gets from (\ref{Ge})
\begin{equation}
 <\hat n_\sigma> = (1-<\hat n_{\bar\sigma}>) I_{2,\sigma}
                 + <\hat n_{\bar\sigma}> I_{3,\sigma}
\end{equation}
and hence
\begin{equation}
 \label{nsigma}
 <\hat n_\sigma> = 
 \frac{I_{2,\sigma}-I_{2,\bar\sigma}[I_{2,\sigma}-I_{3,\sigma}]}
      {1-[I_{2,\bar\sigma}-I_{3,\bar\sigma}][I_{2,\sigma}-I_{3,\sigma}]}
\end{equation}
$G^{(e)<}_{i,\sigma}$ can be calculated from the Keldysh equation
\begin{equation}
 G^{(e)<}_{i,\sigma}(E) = 
 G^{(e)r}_{i,\sigma}(E)\,\Sigma^{(e)<}_{i,\sigma}(E)\,G^{(e)a}_{i,\sigma}(E)
\end{equation}
with $\Sigma^{(e)<}_{i,\sigma}$ ($i=1,2,3,4$) being lesser projections of the
corresponding self-energies presented in Eqs.(\ref{Ge1_m1})-(\ref{Ge4_m1}),
i.e.
\begin{align}
 \Sigma^{(e)}_{1,\sigma}(\tau,\tau') &= 
 \Sigma_{\sigma 0}(\tau,\tau')+\Sigma_{\sigma 3}(\tau,\tau')
 \\
 \Sigma^{(e)}_{2,\sigma}(\tau,\tau') &=
 \Sigma_{\sigma 0}(\tau,\tau') - U\int_c d\tau''\,
      G_{1,\sigma}^{(e)}(\tau,\tau'')\Sigma_{\sigma 1}(\tau'',\tau')
 \\
 \Sigma^{(e)}_{3,\sigma}(\tau,\tau') &=
 \Sigma_{\sigma 0}(\tau,\tau')+ U\int_c d\tau''\,
      G_{4,\sigma}^{(e)}(\tau,\tau'')\Sigma_{\sigma 2}(\tau'',\tau')
 \\
 \Sigma^{(e)}_{4,\sigma}(\tau,\tau') &=
 \Sigma_{\sigma 0}(\tau,\tau')+\Sigma_{\sigma 3}(\tau,\tau')
\end{align}
and expressions for $\Sigma_{\sigma i}$ ($i=\{0,1,2,3\}$)
given by (\ref{SEs0})-(\ref{SEs3}).

Since $G^{(e)<}_{i,\sigma}$ ($i=1,2,3,4$) and therefore $I_{i,\sigma}$
do not depend on $<\hat n_{\sigma}>$, Eq.(\ref{nsigma}) is an explicit
expression for $<\hat n_{\sigma}>$ and not, as might have expected,
an equation that needs to be solved self-consistently.
Eq.(\ref{Ge}) therefore constitutes an explicit expression for $G^{(e)}_\sigma$ 
that can be evaluated directly once the $G^{(e)<}_{i,\sigma}$ are known.
Thus the Keldysh contour based consideration
provides full information on the nonequilibrium system, and no separate
considerations (as non-crossing approximation used in 
Ref.~\onlinecite{MeirWingreenLee_Kondo}) are needed in order to estimate 
the level population. Note that both 
Ref.~\onlinecite{MeirWingreenLee_Kondo} and our consideration 
give only qualitative description of the Kondo effect,
since correlation between localized spin at the level and 
opposite spin cloud in the contacts is treated perturbatively.

When electron-phonon interaction is present Eq.(\ref{defnsigma}) 
remains valid. This results from the fact that $K^{<}(t,t)=1$
so that $G_{\sigma}^{<}(t,t)=G_{\sigma}^{(e)<}(t,t)$; still,
one has to deal with a self-consistent procedure. Indeed, the phonon GF
$D_{P_aP_a}$ (and hence shift generator correlation function $K$, 
see Eq.~(\ref{XXKeldysh})) depends on the electronic GF $G_\sigma^{(e)}$ 
through its `self-energy' $\Pi_{P_aP_a}$, Eq.~(\ref{DSEKeldysh}). 
On the other hand, the electron GF $G_\sigma^{(e)}$ depends on the
shift generator correlation function $K$ through its `self-energies'
$\Sigma_{\sigma i}$ ($i=\{0,1,2,3\}$), Eqs.(\ref{SEs0})-(\ref{SEs3}).
The resulting procedure is described in detail in Ref.~\onlinecite{strong_elph}.
The only difference that enters here is the need to obtain the different
self-energies defined in Eqs.(\ref{SEs0})-(\ref{SEs3}).

As discussed in Ref.~\onlinecite{strong_elph} the calculations involving
electron-phonon interaction, when multiplication by the FC factor is 
necessary, are facilitated by repeatedly moving between the time and energy 
domains. This is done using fast Fourier transform (FFT).
In the calculations we use (following \cite{WingreenMeir})
for the retarded projection of $\Sigma_{K,\sigma 0}$
\begin{equation}
 \label{SigmaK_E}
 \Sigma_{K,\sigma 0}^{(e)\, r}(E)=
 \frac{1}{2}
 \frac{\Gamma_{K,\sigma}^{(0)}W_{K,\sigma}^{(0)}}
 {E-E_{K,\sigma}^{(0)}+iW_{K,\sigma}^{(0)}}
\end{equation}
while its lesser projection is given by (\ref{SEKlt}), where
\begin{equation}
 \label{GammaK_E}
 \Gamma_{K,\sigma 0}(E) = 
 -2\mbox{Im}\left[\Sigma_{K,\sigma 0}^{(e)\, r}(E)\right]
\end{equation}
We take $W_{K,\sigma}^{(0)}=10 U$ and $E_{K,\sigma}^{(0)}$ taken at the 
Fermi level, defined to be the zero of energy ($E_F=0$). 
This form will ensure convergence of the integrals.
A band width ten times the Coulomb repulsion is enough to get essentially
constant density of contacts states in the relevant energy region 
(wide band).
$\Gamma_{K,\sigma}^{(0)}$ is taken much smaller than $U$ to simulate
the Coulomb blockade regime; exact numbers are indicated in calculation 
parameters below. 

The biased junction was characterized by the choice
\begin{equation}
 \mu_L=E_F+\eta\,eV_{sd} \qquad \mu_R=E_F-(1-\eta)\,eV_{sd}
\end{equation}
with voltage division factor $\eta=0.5$. 
In calculations with $M\neq 0$, where an iterative
procedure was used, convergence was assumed when population differences 
(electronic population for both spins and vibrational population) 
between consecutive iteration steps were less than predefined tolerance, 
taken to be $10^{-4}$. The application of a gate potential was represented
by taking 
\begin{equation}
 \label{Vg}
 \bar\varepsilon_\sigma(V_g)=\bar\varepsilon_\sigma(V_g=0)+eV_g
\end{equation}
Note that $V_g$ in (\ref{Vg}) is the effective potential at the molecule,
which is usually considerably smaller than the bare potential applied
to the gate.

In what follows we apply the procedure outlined above in two 
situations. In section~\ref{results} we focus on Coulomb blockade
phenomena.
In section~\ref{kondo} we describe the application
to Kondo physics by keeping the temperature
low enough and by assigning finite lifetimes to the metal electrons.

\section{\label{results}Numerical results in the Coulomb Blockade regime}
When dealing with the Coulomb blockade type calculations,
the electronic part (without the Franck-Condon (FC) factors) 
of the lesser and greater projections of $\Sigma_{\sigma j}$ ($j=1,2,3$)
are obtained from Eqs.(\ref{SEs1})-(\ref{SEs3}) and given by
\begin{align}
 \label{SEs1lt}
 &\Sigma_{\sigma 1}^{(e)\, <}(E) = 
 \\&
 i\sum_{K=L,R}
 \left[\Gamma_{K,\bar\sigma}(E_{1\sigma})f_K^2(E_{1\sigma})
      +\Gamma_{K,\bar\sigma}(E_{2\sigma})f_K^2(E_{2\sigma})\right]
 \nonumber\\
 \label{SEs1gt}
 &\Sigma_{\sigma 1}^{(e)\, >}(E) = 
 \nonumber\\&
 -i\sum_{K=L,R}
 \left[\Gamma_{K,\bar\sigma}(E_{1\sigma})f_K(E_{1\sigma})[1-f_K(E_{1\sigma})]
 \right.\\ &\qquad\qquad \left.
      +\Gamma_{K,\bar\sigma}(E_{2\sigma})f_K(E_{2\sigma})[1-f_K(E_{2\sigma})]
 \right]
 \nonumber\\
 \label{SEs2ltgt}
 &\Sigma_{\sigma 2}^{(e)\, >,<}(E) = \Sigma_{\sigma 3}^{(e)\, >,<}(E)
                                   - \Sigma_{\sigma 1}^{(e)\, >,<}(E)
 \\
 \label{SEs3lt}
 &\Sigma_{\sigma 3}^{(e)\, <}(E) = 
 \nonumber\\&
 i\sum_{K=L,R}
 \left[\Gamma_{K,\bar\sigma}(E_{1\sigma})f_K(E_{1\sigma})
      +\Gamma_{K,\bar\sigma}(E_{2\sigma})f_K(E_{2\sigma})\right]
 \\
 \label{SEs3r}
 &\Sigma_{\sigma 3}^{(e)\, >}(E) = 
 -i\sum_{K=L,R}
 \left[\Gamma_{K,\bar\sigma}(E_{1\sigma})[1-f_K(E_{1\sigma})]
 \right.\nonumber\\&\left. \quad\qquad\qquad\qquad\qquad
      +\Gamma_{K,\bar\sigma}(E_{2\sigma})[1-f_K(E_{2\sigma})]\right]
\end{align}
where $E_{1\sigma}=\bar\varepsilon_\sigma+\bar\varepsilon_{\bar\sigma}+U-E$
and $E_{2\sigma}=E-\bar\varepsilon_\sigma+\bar\varepsilon_{\bar\sigma}$.
Retarded projection of the full SEs (after dressing by FC factors)
are obtained using Lehmann representation.\cite{Mahan}

\begin{figure}[htbp]
\centering\includegraphics[width=\linewidth]{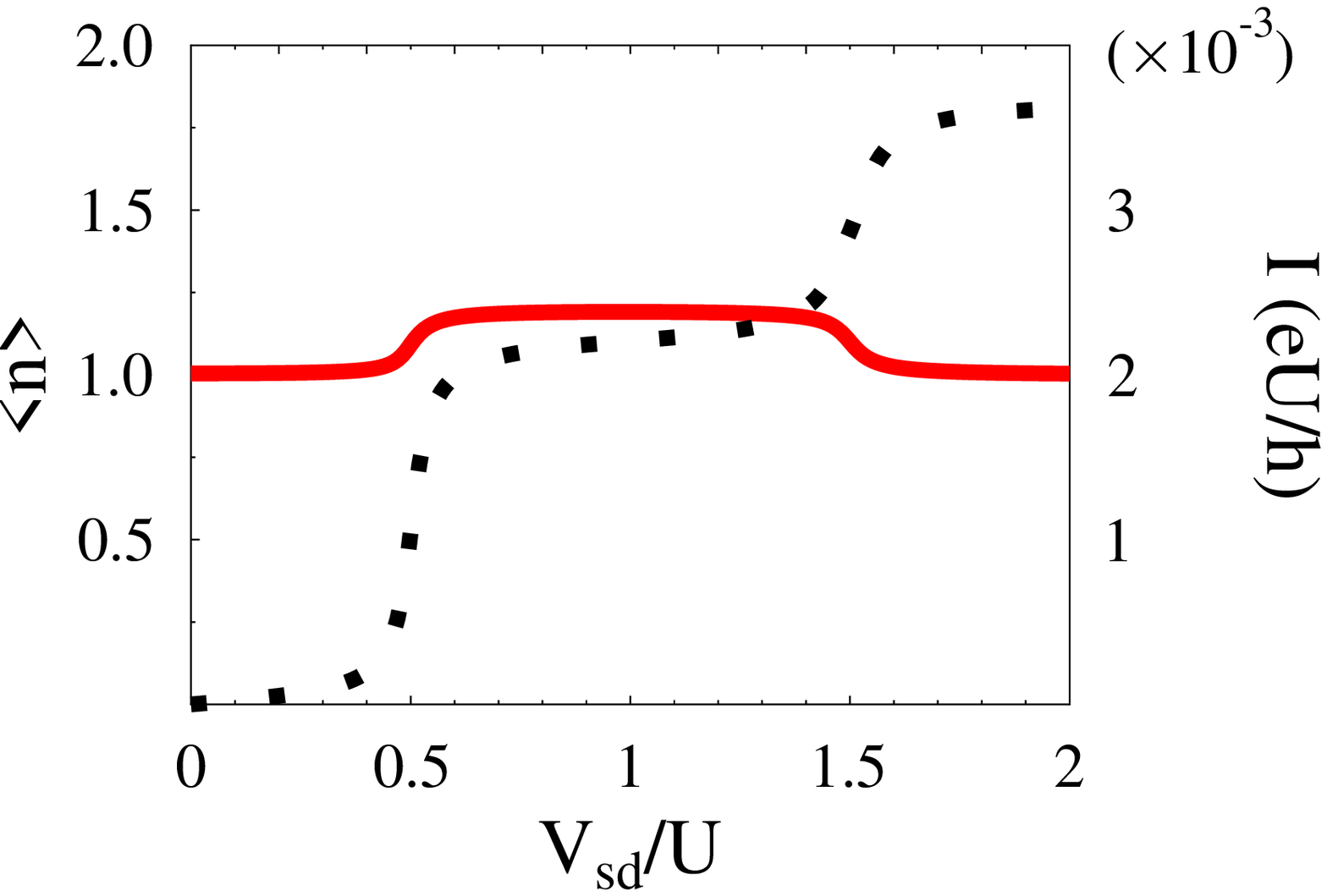}\\
\centering\includegraphics[width=\linewidth]{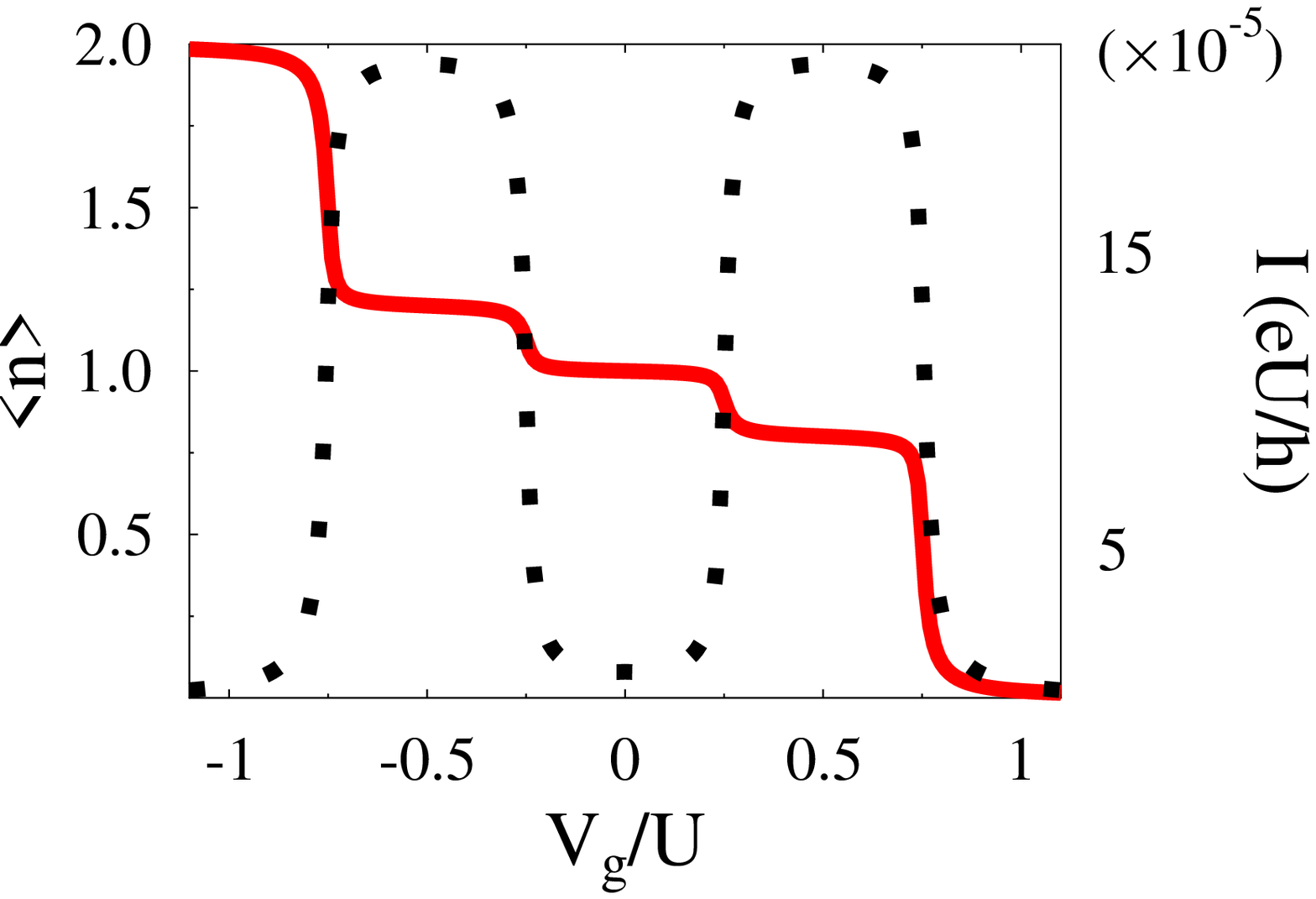}\\
\centering\includegraphics[width=\linewidth]{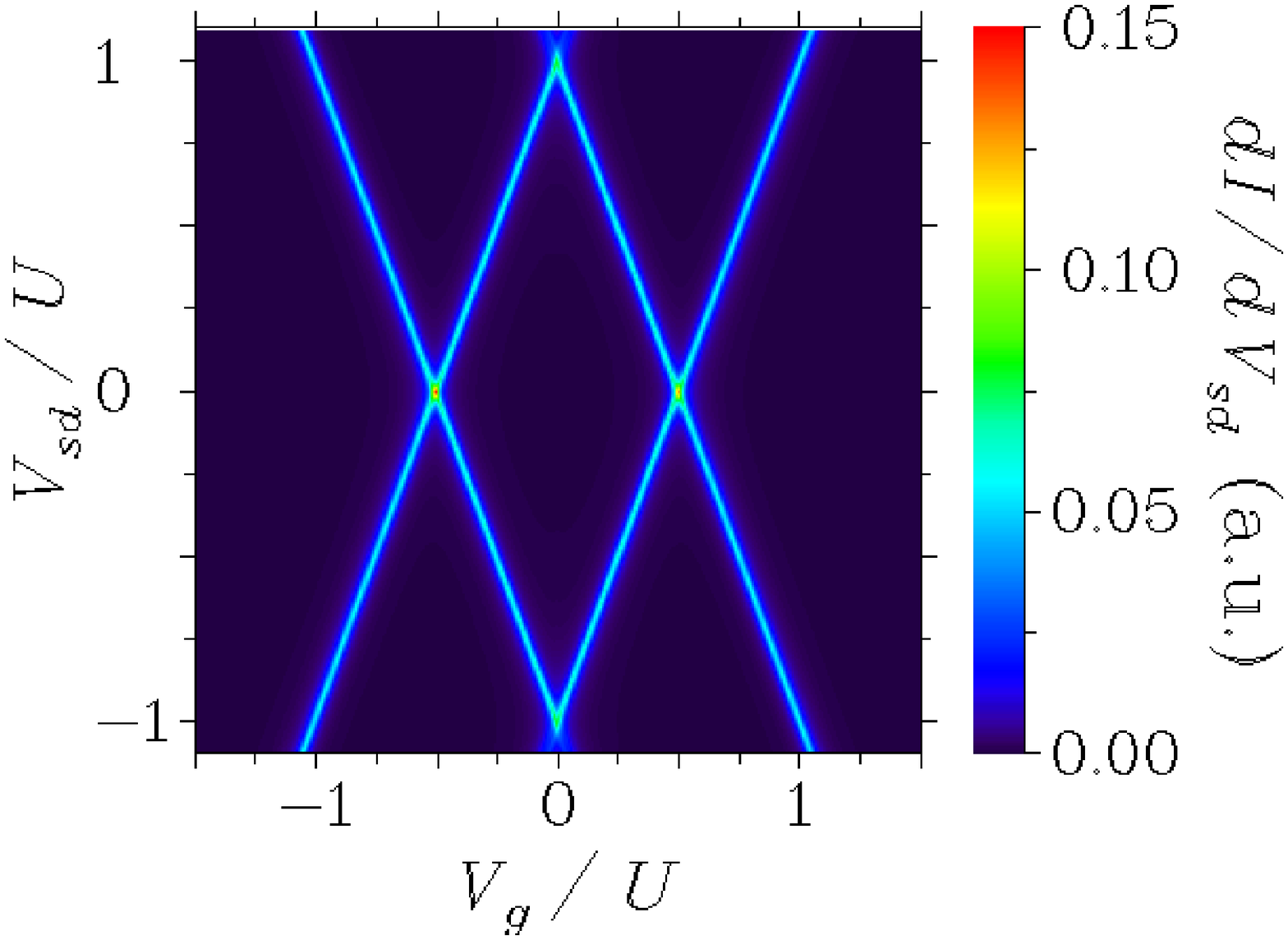}
\caption{\label{fig_hub}
(Color online) Elastic resonant tunneling. 
Average population (solid line, red; left axis) and 
current (dotted line, black; right axis) as function of  
(a) $V_{sd}$ at fixed $V_g=-U/4$ and (b) $V_g$ at fixed $V_{sd}=U/2$.
(c) Contour plot of $dI/dV_{sd}$ vs. $V_g$ and $V_{sd}$. 
See text for parameters. Note that $V_{sd}$ axis range in (a) goes
beyond that in (c).
}
\end{figure}

Consider first the situation where
no electron-phonon coupling is present, $M=0$. Figure~\ref{fig_hub}c 
shows a conductance contour plot as a function of the gate and 
source-drain voltages for a system characterized by
$\varepsilon_\sigma=-0.5$, $\Gamma_{K,\sigma}^{(0)}=0.01$, and $T=10^{-4}$
(all parameters are in units of $U$).
Fig.~\ref{fig_hub}a presents average level population (solid line) and
current (dotted line) plotted as a function of $V_{sd}$ at fixed $V_g=-U/4$.
$I/V_{sd}$ curve shows two Coulomb addition plateaus, as is expected
for a doubly degenerate single level.
Fig.~\ref{fig_hub}b is a similar graph as function of $V_g$
at fixed $V_{sd}=U/2$. The usual Coulomb blockade diamond structure
is observed in the bottom graph.  Naturally, at high positive $V_g$
the level is unpopulated, while at high negative $V_g$ it is fully populated 
($<\hat n>=2$). Within the conduction diamond the average population is 1, 
indicating the Coulomb blockade situation. Intermediate regions provide 
fractional average populations due to partial occupation of the levels. 

\begin{figure}[htbp]
\centering\includegraphics[width=\linewidth]{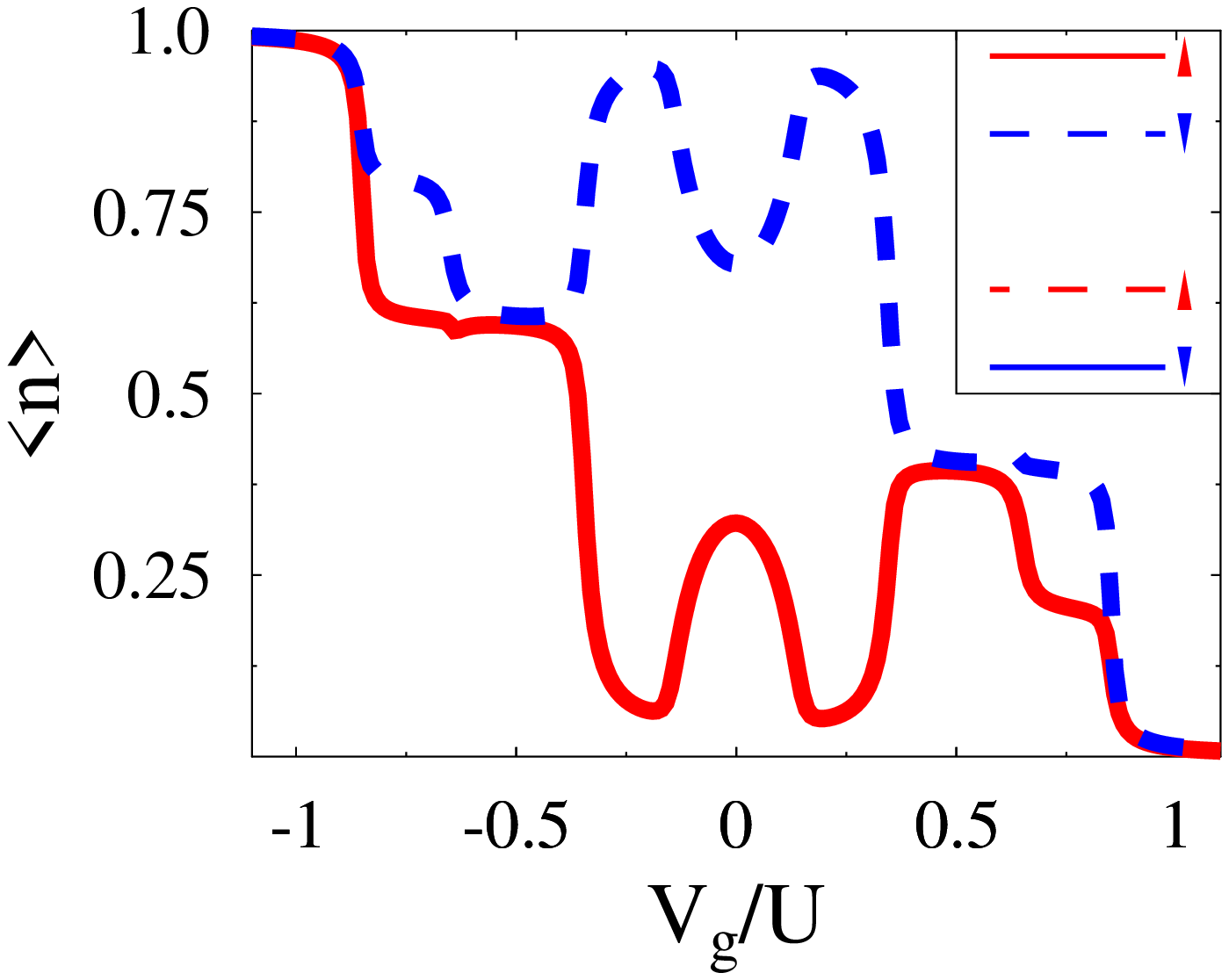}\\
\centering\includegraphics[width=\linewidth]{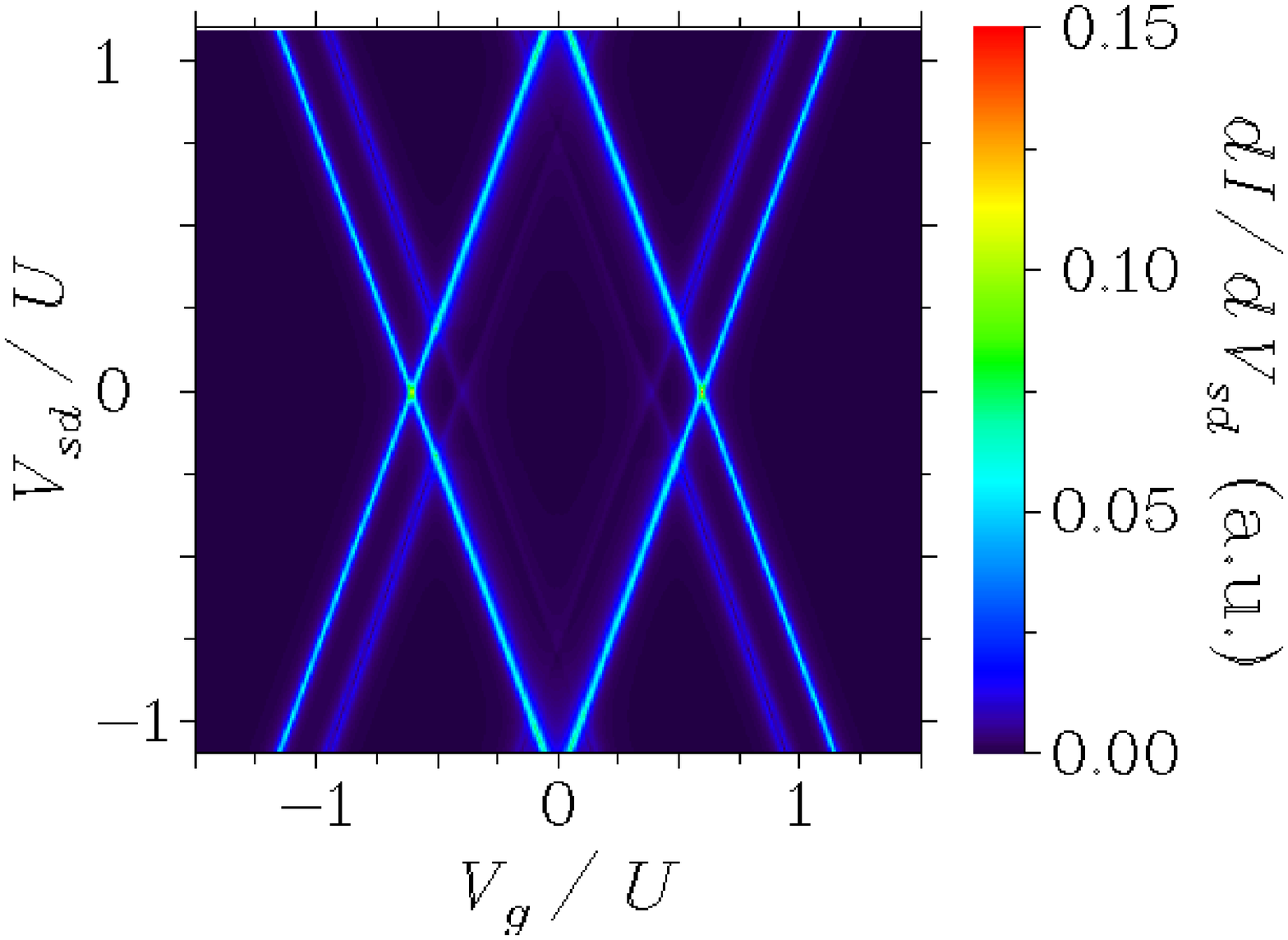}
\caption{\label{fig_hub_b}
(Color online) Elastic resonant tunneling under applied magnetic field. 
Average populations of spin up (solid line, red) and 
spin down (dashed line, blue) levels vs. $V_g$ at fixed $V_{sd}=U/2$ (top).
Contour plot of $dI/dV_{sd}$ (bottom) vs. $V_g$ and $V_{sd}$. 
Inset shows two possible states (solid and dashed lines) 
of the molecule.  See text for parameters.
}
\end{figure}

The case $\varepsilon_\sigma\neq\varepsilon_{\bar\sigma}$,
that may correspond to magnetic field removal of spin degeneracy
is shown in Figure~\ref{fig_hub_b}.
We take the split levels to be $\varepsilon_{\downarrow}=-0.6$
and $\varepsilon_{\uparrow}=-0.4$, other parameters are identical to those of
Fig.~\ref{fig_hub}. This split results in
splitting of the conductance lines as is shown in the bottom graph. 
Note the different intensity of the lines outside the diamond,
The difference becomes even more drastic inside the diamond. This result
is in agreement with experimental observation.\cite{McEuen} 
The calculated average population of the two spin levels (top graph),
where again the source-drain voltage is fixed at $V_{sd}=U/2$, 
shows their complex dependence on gate voltage.
This behavior can be understood within a simple argument.
The molecule in the junction can be in either of the two states
sketched in the inset of the top graph by solid and dashed lines.
The observed average is the sum of the two contributions with weights 
representing
probability for the system to be in the state. A qualitative explanation is
based on the assumption that the system strives to be in a minimum energy
situation (note that this explanation is only qualitative, since an energy 
minimum is not required in the nonequilibrium transport case, however it might work
to some extent in the blockade regime). Thus the probability to be in the
state indicated by solid lines in the inset is much higher than in the other.
So, the most pronounced lines in conductance appear when chemical potentials 
cross the energy levels of this (solid line levels in the inset) state.
Average population behavior can be explained with this consideration as well.

\begin{figure}[htbp]
\centering\includegraphics[width=\linewidth]{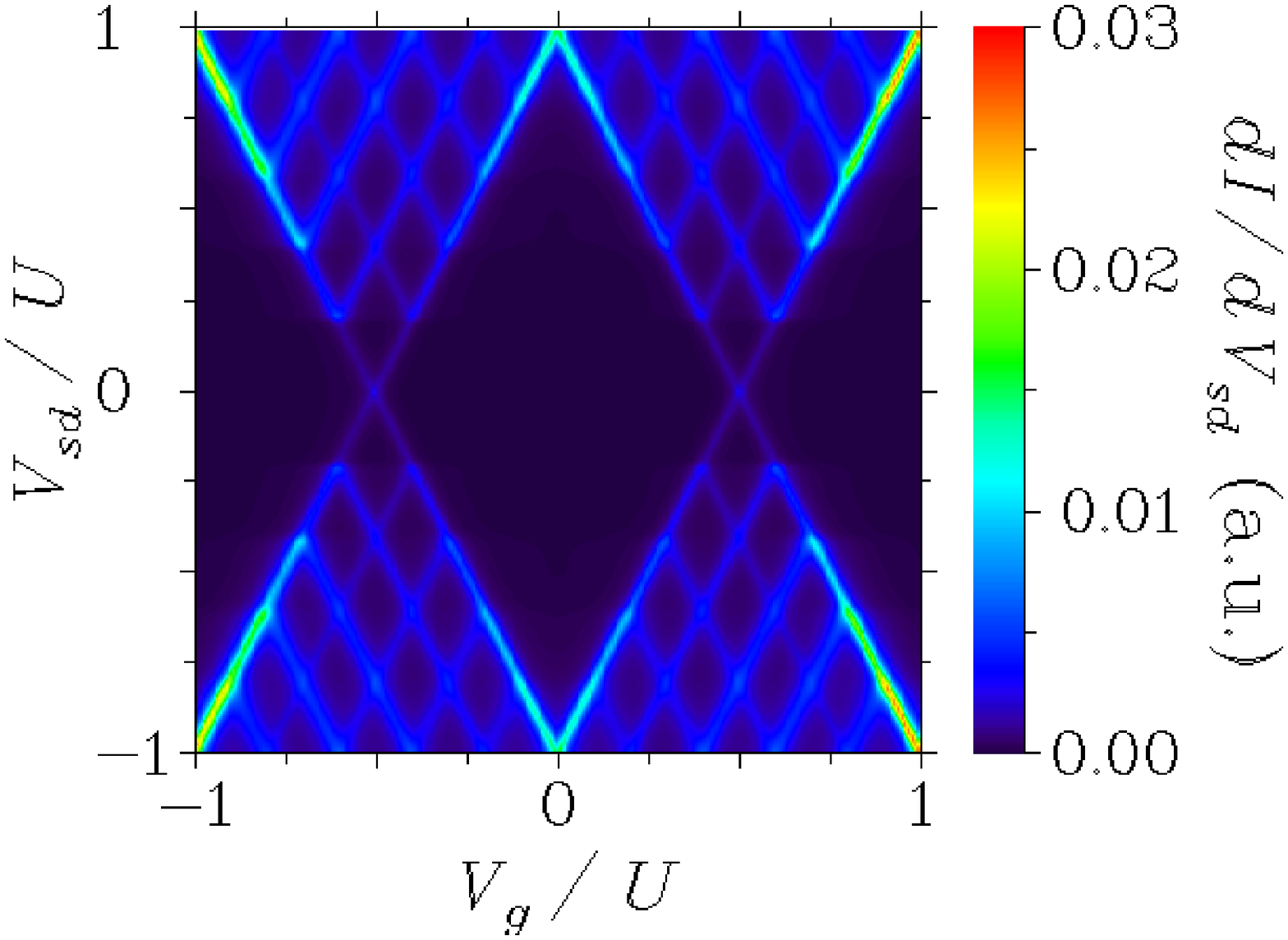}\\
\centering\includegraphics[width=\linewidth]{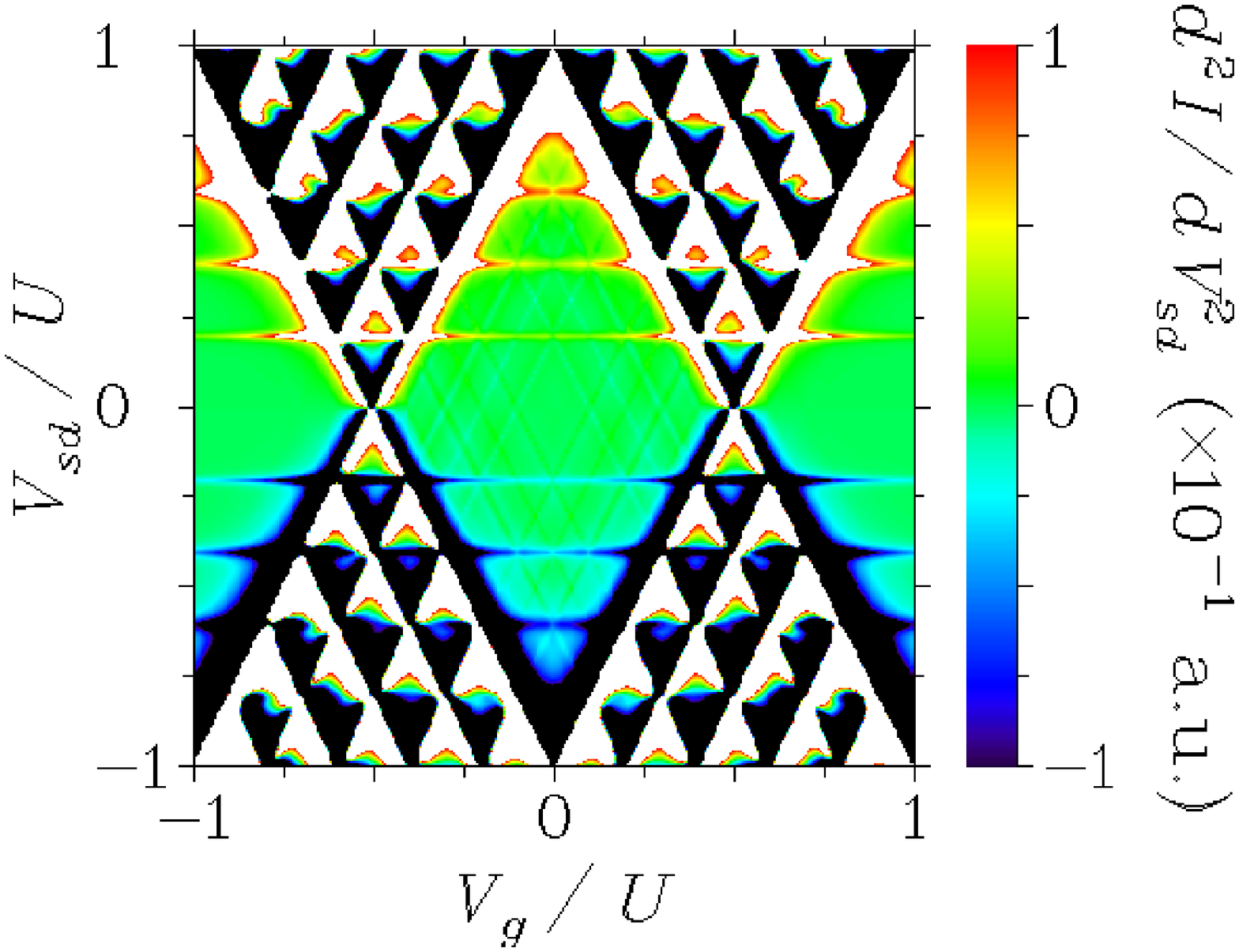}\\
\centering\includegraphics[width=\linewidth]{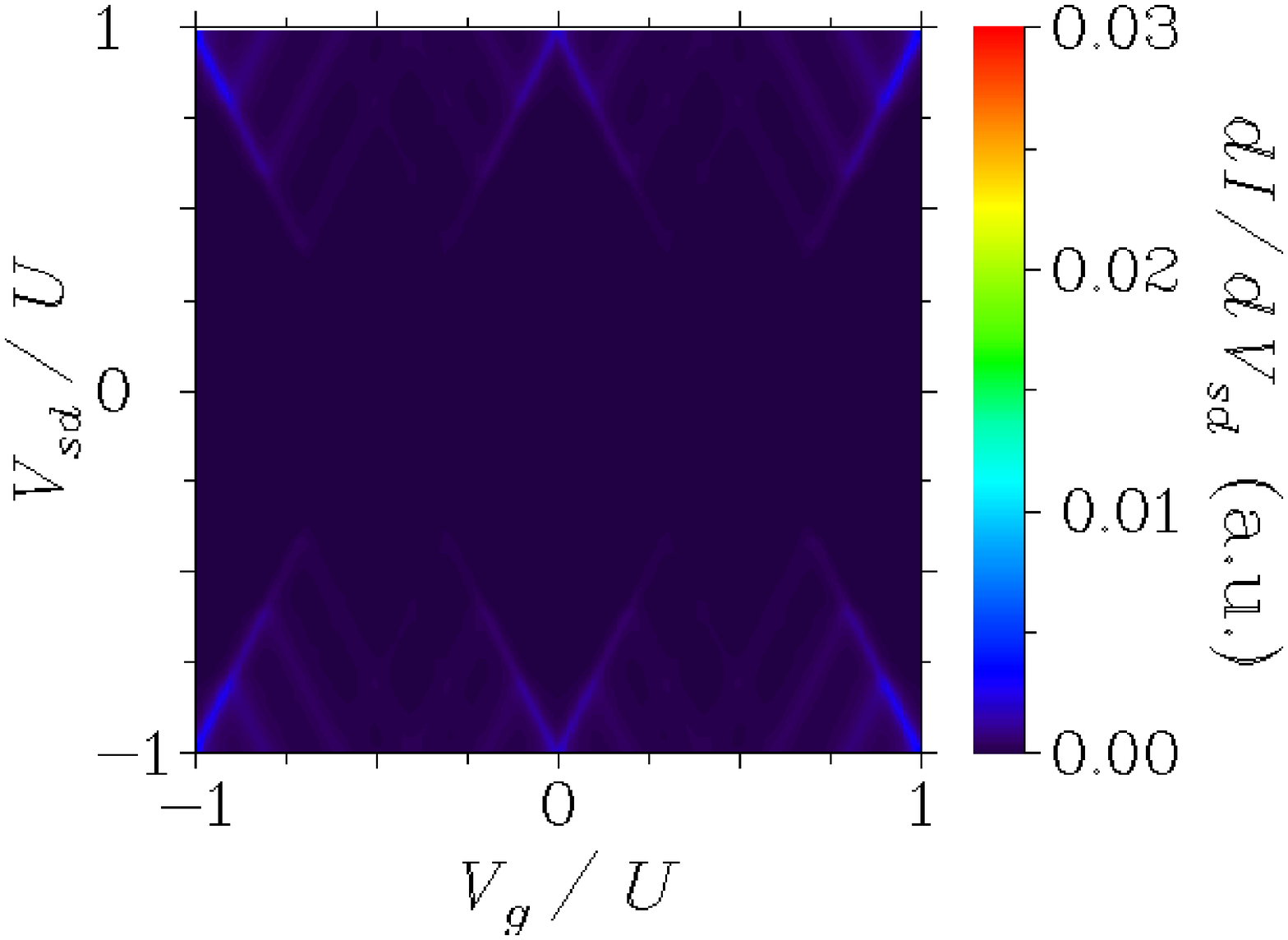}
\caption{\label{fig_lam}
(Color online) Inelastic tunneling. 
(a) Contour plot of $dI/dV_{sd}$ vs. $V_g$ and $V_{sd}$. 
(b) Contour plot of $d^2I/dV_{sd}^2$ vs. $V_g$ and $V_{sd}$. 
White regions correspond to values outside the scale;
(c) Franck-Condon blockade.
See text for parameters.
}
\end{figure}

In the presence of vibrational degrees of freedom inelastic co-tunneling
(vibrational inelasticity) can be observed in 
conductance.\cite{Natelson,Zant,Natelson_review}
The situation is illustrated within a  zero-order 
calculation\cite{zero_order} using the parameters (in units of $\bar U$)
$T=10^{-3}$, $\bar\varepsilon_\sigma=-0.5$, $\Gamma_{K,\sigma}^{(0)}=0.01$,
$\omega_0=0.2$, and $M=0.4$.
The following points should be noted:
\begin{enumerate}
\item Figure~\ref{fig_lam}a shows the main Coulomb steps in the conductance 
 map. 
 In addition to elastic, vibrational sidebands corresponding
 to phonon creation by the tunneling electron are observed.
 Peaks corresponding to phonon absorption are not seen due to 
 the low temperature employed in the calculation.
\item Figure~\ref{fig_lam}b represents the second derivative of current 
 vs. source-drain voltage map. In addition to resonant 
 vibrational sidebands 
 (lines along main Coulomb steps) observed in Fig.~\ref{fig_lam}a 
 here one sees also inelastic electron tunneling spectroscopy (IETS)
 vibrational features
 (gate voltage independent off-resonant vibrational features)
 as well as weak lines corresponding to phonon annihilation.
\item The absence of vibrational sidebands fro variable $V_g$
 for $V_{sd}<\omega_0$ is clearly seen from Fig.~\ref{fig_lam}a. 
 This issue was first addressed in Ref.~\onlinecite{Mitra} and later confirmed 
 by us.\cite{strong_elph}
\item Suppression of the conduction signal at low source-drain voltage
 (the so called Franck-Condon blockade \cite{KochvonOppen})
 is seen from Fig.~\ref{fig_lam}a as well.
 At even stronger electron-phonon coupling (Figure~\ref{fig_lam}c; a zero-order
 calculation with the same parameters as in Fig.~\ref{fig_lam}a except that
 $M=0.6$), the low voltage signal is suppressed completely.
\item Note, that while experimentally the scales in $V_g$ and $V_{sd}$ where
Coulomb blockade diamonds are observed are very different
($V_{sd}$ is of order of Coulomb repulsion energy, $100$~mV, 
while $V_g$ spans $\sim 1$~V), in our 
calculations they are comparable. The reason for this is that experimentally 
only part of the applied gate voltage affects the position of 
the molecular level relative to contact Fermi energy. This is due to two 
reasons: first, capacitance factors (charging of the junction) play a role, 
and second, gate voltage can not be tuned to strongly affect the molecule
because of small sizes of the junction.\cite{GhoshDatta_gate}
In our calculations however a rigid shift of molecular level is assumed.
\end{enumerate}

\section{\label{kondo}The Kondo regime}
The Kondo effect\cite{Hewsonbook}, a crossover from weak to strong coupling 
between localized (molecular) and band (contacts) electrons, manifests itself 
in molecular junctions as a maximum in electrical conductance near
$V_{sd}\sim 0$ at low temperatures.
Conduction in this regime was described by 
Meir~et~al.\cite{MeirWingreenLee_Kondo} within an EOM scheme.
The treatment has focused on the retarded GFs, making it necessary
to get level populations from a separate calculation using the
non-crossing approximation (NCA). 
In contrast, the NEGF EOM approach yields both the retarded and lesser GFs,
and the needed level populations are obtained from the latter. This provides 
a single consistent theoretical framework that, as we show below, 
reproduces the results of Ref.~\onlinecite{MeirWingreenLee_Kondo}.
It should be noted however that this approach is still an approximation,
since truncating the EOM hierarchy, Eqs.(\ref{defGam3e1})-(\ref{defGam3e6}), 
implies neglect
of correlations that may become important in the mixed valence situation
when the level $\varepsilon_\sigma$ (shifted by $V_g$) is close to the
Fermi energy. Therefore our nonequilibrium treatment of the Kondo regime
is questionable beyond the low bias regime $E_F-\varepsilon_\sigma\gg eV_{sd}$,
similar to the mean-field slave boson
approach\cite{Hewsonbook,Guo,AguadoLangreth,KangChoKimShin}
where charge correlations are neglected by the mean-field approximation.
In both approaches though the needed correlations in spin fluctuations
are maintained; in the present approach this is done by keeping the
correlation functions (\ref{defGam2e1})-(\ref{defGam2e3}) as essential 
ingredients of the calculation.

Consider first the purely electronic case, $M=0$.
Following \cite{MeirWingreenLee_Kondo} we limit our consideration to the
$U\to\infty$ limit. This leads to significant simplification while at the
same time limiting the site to at most single occupancy as required for 
observation of the Kondo effect.\cite{even_Kondo} 
From Eqs.~(\ref{Gei})-(\ref{Ge4_m1})
it follows that $G_{3,\sigma}^{(e)}\sim 1/U\to 0$ in this limit, while
$G_{2,\sigma}^{(e)}\to G_{2,\sigma}^{(e,\infty)}$ 
satisfies the following Dyson equation
\begin{align}
  \label{G2seinfty}
  &\int_c d\tau\, \left[
  \left(i\frac{\partial}{\partial\tau}-\varepsilon_\sigma\right)
  \delta(\tau_1,\tau)-\Sigma_{\sigma 0}(\tau_1,\tau)
 -\Sigma_{\sigma 1}^{(\infty)}(\tau_1,\tau) \right]
 \nonumber\\&\qquad\times
  G_{2,\sigma}^{(e,\infty)}(\tau,\tau_2) = \delta(\tau_1,\tau_2)
\end{align}
with $\Sigma_{\sigma 0}$ defined in (\ref{SEs0}) and
from Eq.(\ref{SEs1}) (because $g_{k,\bar\sigma}^{(1)}\to 0$ in the
$U\to\infty$ limit; c.f. Eq.(\ref{defg1}))
\begin{equation}
 \label{SEs1infty}
 \Sigma_{\sigma 1}^{(\infty)}(\tau,\tau') = 
 \sum_{K=L,R}\sum_{k\in K}\left|V_{k\bar\sigma}\right|^2
 \left<\hat n_{k\bar\sigma}\right> g_{k,\bar\sigma}^{(2)}(\tau,\tau')
\end{equation}
Thus from (\ref{Ge}) it follows that the total GF in the $U\to\infty$ limit is
\begin{equation}
 \label{Geinfty}
 G_\sigma^{(e,\infty)} = \left[1-<\hat n_{\bar\sigma}>\right] 
 G_{2,\sigma}^{(e,\infty)}(\tau_1,\tau_2)
\end{equation}
The Kondo peak diverges unless the finite lifetime of metal electrons is
taken into account. We incorporate this lifetime in the form introduced in
Eq.(5) of Ref.~\onlinecite{MeirWingreenLee_Kondo} (which associates lifetime with
scattering off the molecular state).
Note that the Lorentzian form adopted following\cite{WingreenMeir} 
for the coupling between molecule and contacts, Eq.~(\ref{SigmaK_E}), 
prevents ultraviolet divergence of integrals such as (\ref{SEs1rE})
and allows analytic evaluation of $\Sigma_{\sigma 1}^{(\infty)}$ projections 
(see Appendix~\ref{B}, Eqs.~(\ref{SEs1r_infty}) and (\ref{SEs1lt_infty})).

\begin{figure}[htbp]
\centering\includegraphics[width=\linewidth]{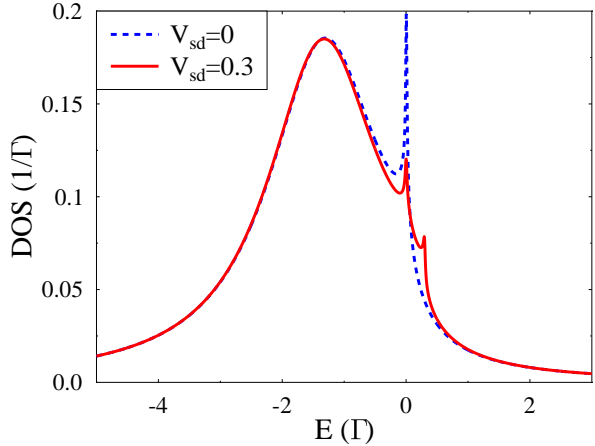}
\caption{\label{fig_kondo_reproduce}(Color online)
Bridge density of states in the Kondo regime for equilibrium 
(dashed line, blue) and nonequilibrium (solid line, red) situations.
See text for parameters.
}
\end{figure}

Eqs.~(\ref{G2seinfty})-(\ref{Geinfty}) lead to the following form for the
retarded projection of $G_{\sigma}^{(e,\infty)}$
\begin{equation}
 \label{Gseinfty_r}
 G_{2,\sigma}^{(e,\infty)r}(E)=\frac{1-<\hat n_{\bar\sigma}>}
 {E-\varepsilon_\sigma-\Sigma_{\sigma 0}(E)-\Sigma_{\sigma 1}^{(\infty)}(E)}
\end{equation}
where $\Sigma_{\sigma 0}(E)$ and $\Sigma_{\sigma 1}^{(\infty)}(E)$ 
are defined in in Eqs.~(\ref{SigmaK_E}) and (\ref{SEs1r_infty}) respectively.
These expressions are identical to Eqs.~(3) and(4) of 
Ref.~\onlinecite{MeirWingreenLee_Kondo}. 
Note however that $<\hat n_{\bar\sigma}>$
is now calculated from the lesser projection 
\begin{equation}
  \label{Gseinfty_lt}
  G_{\bar\sigma}^{(e,\infty)<}(E) = \left[1-<\hat n_{\bar\sigma}>\right]
  G_{2,\sigma}^{(e,\infty)<}(E)
\end{equation}
Figure~\ref{fig_kondo_reproduce} presents the bridge density of states in 
equilibrium (dashed line) and nonequilibrium (solid line) situations.  
Parameters of the calculation are (in units of 
$\Gamma_\sigma^{(0)}=\Gamma_{L,\sigma}^{(0)}+\Gamma_{R,\sigma}^{(0)}$) 
$T=0.005$, $\varepsilon_\uparrow=\varepsilon_\downarrow=-2$,
$W_{K,\sigma}^{(0)}=100$.
As before the equilibrium Fermi energy defines the energy origin, and the
nonequilibrium situation is characterized by $\mu_L=E_F+|eV|$ and $\mu_R=E_F$.
In equilibrium a Kondo peak at the Fermi energy is seen. It splits into two
(at each of the electrode Fermi energies) when finite bias is applied. 
Comparing to Figs.~1a and b of Ref.~\onlinecite{MeirWingreenLee_Kondo}
we see that the present formalism essentially reproduces these results.

\begin{figure}[htbp]
\centering\includegraphics[width=\linewidth]{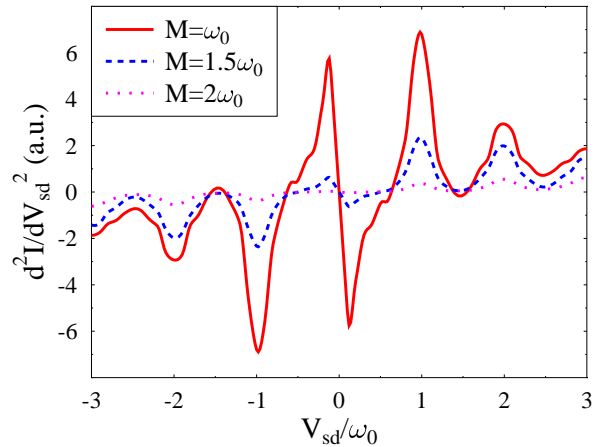}
\caption{\label{fig_kondo_ph}(Color online)
$d^2I/dV_{sd}^2$ for molecular junction in the Kondo regime. Shown are results
for three choices of strength in coupling to vibration:
$M=\Gamma/2$ (solid line, red), $M=2\Gamma/3$ (dashed line, blue), 
and $M=\Gamma$ (dotted line, magenta). See text for other parameters.
}
\end{figure}

Inelastic effects are introduced into the picture as before, by 
dressing transfer matrix elements by the shift operators, see 
Eqs.~(\ref{appGFKeldysh}), (\ref{SEs0}) and (\ref{SEs1}).
Figure~\ref{fig_kondo_ph} shows the result, obtained from such calculation
for the second derivative of the current 
with respect to the source-drain voltage (bottom graph), for three choices
of the electron-vibration coupling strength. Parameters of the calculation
are (in units of $\Gamma_\sigma^{(0)}$) $T=0.025$, $\varepsilon_\sigma=-2$,
$W_{K,\sigma}^{(0)}=100$, $\omega_0=0.5$. The solid, dashed, and dotted
lines correspond to $M=0.5$, $0.75$, and $1$, respectively. 
As is expected, increase
in electron-vibration interaction destroys the Kondo effect. 
The reasons for this are (a) dephasing due to 
electron-vibration interaction and (b) shift of the energy level due to
phonon reorganization. 
Electronic level shift downwards decreases the Kondo temperature
($T_K\sim\exp{-\pi|\varepsilon_\sigma|/\Gamma_\sigma}$ see 
Ref.~\onlinecite{Hewsonbook}) thus destroying the Kondo peak.

It should be emphasized that the vibrational structure seen in 
Fig.~\ref{fig_kondo_ph} is a normal inelastic tunneling feature that
is seen to persist also in the Kondo regime. This feature appears 
both in the Kondo and in the normal blockade regimes 
(see Figs.~\ref{fig_lam}b and \ref{fig_kondo_ph}), as indeed
was recently observed in the molecular junction experiment of 
Yu~et~al.\cite{Natelson} The transition between these regimes 
(when a molecular orbital crosses the Fermi energy) can not be
described by our approach for reasons outlined above.
Also, Paaske and Flensberg\cite{Flensberg} have recently applied 
a perturbative renormalization group to a limiting form of the same model
in which the molecular electronic level is always in equilibrium 
with one side of the junction (the substrate in an STM configuration)
and have shown that maintaining quantum coherence of vibrons,
the effect disregarded in our treatment due to approximation 
(\ref{appGFKeldysh}), may lead to enhancement of the exchange coupling and 
hence the Kondo temperature.

\section{\label{conclude}Conclusion}
We study inelastic effects in electron transport through a model molecular 
junction in Coulomb blockade and Kondo regimes. 
The approach is based on nonequilibrium
generalization of the equation-of-motion scheme introduced by 
Meir~et~al.\cite{MeirWingreenLee_CB,MeirWingreenLee_Kondo} and is appealingly
simple. Inelastic effects are treated within a diabatic Born-Oppenheimer scheme.
Important features of this approach are correct analytical 
results for both isolated molecule (no contacts) and
noninteracting ($U=0$) cases, ability to reproduce results by 
Meir~et~al.\cite{MeirWingreenLee_Kondo} without necessity
of additional considerations to get the level population, 
no necessity for self-consistency to get exact (within the scheme)
results when the electron-vibration interaction is switched off,
and unified treatment of both Coulomb and (to some extent) 
Kondo at nonequilibrium.
The approach is able to reproduce experimental features qualitatively.

Inelastic effects obtained within the model are resonant vibrational 
sidebands in the allowed, and IETS signal in the blockaded, parts 
of the conductance map in $V_g-V_{sd}$ coordinates, 
Franck-Condon blockade of transport for relatively strong electron-vibration 
interaction in the Coulomb blockade regime, and vibrational sidebands of the 
Kondo peak, as well as its quenching for strong vibronic coupling.

Generalization of these considerations to the case
 of a two-site molecular bridge in the junction is straightforward.
 The only problem is the large number of equations needed to be taken into
 account in this case. We postpone such generalization for future
 study.

\begin{acknowledgments}
We are grateful to the MURI/DURINT program, to the NASA/URETI program
and to the NSF/MRSEC program for support of this research.
AN thanks the Israel Science Foundation, the US Israel Binational
Science Foundation and the German Israeli Foundation for financial
support of this research.
\end{acknowledgments}

\appendix
\section{\label{A}Derivation of Eq.~(\ref{Ge})}
Here we derive Eq.~(\ref{Ge}). 
Note that the derivation does not depend on whether $V$ or $\bar V$
(and similarly $U$ or $\bar U$) is used for the system-leads coupling
as long as the shift generator operator $X$ is regarded as a scalar.
We follow the procedure invented by
Meir, Wingreen, and Lee\cite{MeirWingreenLee_CB,HaugJauho} for the
equilibrium situation and generalize it to the Keldysh contour
case, in order to take into account the nonequilibrium nature of molecular 
junction transport. During the derivation we will treat transfer matrix
elements $\bar V_{k\sigma}$, Eq.~(\ref{Vks}), as numbers with the
shift generator operators $\hat X_a$, Eq.~(\ref{Xa}), incorporated into 
them as scalar parameters (a Born-Oppenheimer type approximation). 
However we'll have to keep track of their dependence on time
(or more precisely contour variable) in order to get the
phonon correlation functions $K$ correctly at the end.

We start from EOM for GF $G_\sigma^{(e)}(\tau,\tau')$, Eq.~(\ref{defGe}), 
on the Keldysh contour
\begin{align}
 \label{eomGe}
 \left[i\frac{\partial}{\partial\tau}-\bar\varepsilon_\sigma\right]
 G_\sigma^{(e)}(\tau,\tau') &= \delta(\tau,\tau') 
 + \bar U G_\sigma^{(2e)}(\tau,\tau') 
 \\
 &+ \sum_{k\in\{L,R\}} \bar V_{k\sigma}^{\dagger}(\tau)
   \Gamma_{k,\sigma}^{(1e)}(\tau,\tau')
 \nonumber
\end{align}
new GFs on the r.h.s. have the form
\begin{align}
 \label{defGam1e}
 \Gamma_{k,\sigma}^{(1e)}(\tau,\tau') &=  
 -i <T_c \hat c_{k\sigma}(\tau)\, \hat d_\sigma^\dagger(\tau') >
 \\
 \label{defG2e}
 G_\sigma^{(2e)}(\tau,\tau') &= 
 -i <T_c \hat d_\sigma(\tau) \hat n_{\bar\sigma}(\tau)\,
     \hat d_\sigma^\dagger(\tau') >
\end{align}
Now we write EOMs for these GFs 
\begin{align}
 \label{eomGam1e}
 &\left[i\frac{\partial}{\partial\tau}-\varepsilon_{k\sigma}\right]
 \Gamma_{k,\sigma}^{(1e)}(\tau,\tau') = 
 \bar V_{k\sigma}(\tau) G_\sigma^{(e)}(\tau,\tau')
 \\
 \label{eomG2e}
 &\left[i\frac{\partial}{\partial\tau}-\bar\varepsilon_\sigma-\bar U\right]
 G_\sigma^{(2e)}(\tau,\tau') =
 \delta(\tau,\tau')<\hat n_{\bar\sigma}>
 \\
 &\qquad + \sum_k\left[
    \bar V_{k\sigma}^{\dagger}(\tau)\Gamma^{(2e)}_{1,k,\sigma}(\tau,\tau')
   +\bar V_{k\bar\sigma}(\tau)\Gamma^{(2e)}_{2,k,\sigma}(\tau,\tau')
 \right.\nonumber\\&\qquad\qquad\left.
   -\bar V_{k\bar\sigma}^{\dagger}(\tau)\Gamma^{(2e)}_{3,k,\sigma}(\tau,\tau')
    \right]
 \nonumber
\end{align}
While the EOM (\ref{eomGam1e}) closes the chain of equations 
(its r.h.s. contains only $G_\sigma^{(e)}$),
the EOM for $G_\sigma^{(2e)}$ yields new correlations in its r.h.s. defined by
\begin{align}
 \label{defGam2e1}
 \Gamma^{(2e)}_{1,k,\sigma}(\tau,\tau') &=
 -i <T_c \hat c_{k\sigma}(\tau)\hat n_{\bar\sigma}(\tau)\,
         \hat d_\sigma^\dagger(\tau') >
 \\
 \label{defGam2e2}
 \Gamma^{(2e)}_{2,k,\sigma}(\tau,\tau') &=
 -i <T_c \hat c_{k\bar\sigma}^\dagger(\tau)\hat d_\sigma(\tau)
    \hat d_{\bar\sigma}(\tau)\,\hat d_\sigma^\dagger(\tau') >
 \\
 \label{defGam2e3}
 \Gamma^{(2e)}_{3,k,\sigma}(\tau,\tau') &=
 -i <T_c \hat c_{k\bar\sigma}(\tau)\hat d_{\bar\sigma}^\dagger(\tau)
    \hat d_\sigma(\tau)\,\hat d_\sigma^\dagger(\tau') >
\end{align}
As a last step in the chain of EOMs we follow 
references~\cite{MeirWingreenLee_CB,HaugJauho} by writing
equations for the GFs (\ref{defGam2e1})-(\ref{defGam2e3})
\begin{align}
 \label{eomGam2e1}
 &\left[i\frac{\partial}{\partial\tau}-\varepsilon_{k\sigma}\right]
 \Gamma^{(2e)}_{1,k,\sigma}(\tau,\tau') = 
 \bar V_{k\sigma}(\tau) G_\sigma^{(2e)}(\tau,\tau')
 \\
 &\quad + \sum_{k'}\left[
   \bar V_{k'\bar\sigma}(\tau)\Gamma^{(3e)}_{1,k'k,\sigma}(\tau,\tau')
  -\bar V_{k'\bar\sigma}^{\dagger}(\tau)\Gamma^{(3e)}_{2,k'k,\sigma}(\tau,\tau')
  \right]
 \nonumber \\
 \label{eomGam2e2}
 &\left[i\frac{\partial}{\partial\tau}+\varepsilon_{k\bar\sigma}
  -\bar\varepsilon_\sigma-\bar\varepsilon_{\bar\sigma}-\bar U\right]
 \Gamma^{(2e)}_{2,k,\sigma}(\tau,\tau') =
 \nonumber \\ &\quad
 \bar V_{k\bar\sigma}^{\dagger}(\tau)G_\sigma^{(2e)}(\tau,\tau')
 \\
 &\quad - \sum_{k'}\left[
   \bar V_{k'\sigma}^{\dagger}(\tau)\Gamma^{(3e)}_{3,k'k,\sigma}(\tau,\tau')
  +\bar V_{k'\bar\sigma}^{\dagger}(\tau)\Gamma^{(3e)}_{4,k'k,\sigma}(\tau,\tau')
  \right]
 \nonumber \\
 \label{eomGam2e3}
 &\left[i\frac{\partial}{\partial\tau}-\varepsilon_{k\bar\sigma}
 -\bar\varepsilon_\sigma+\bar\varepsilon_{\bar\sigma}\right]
 \Gamma^{(2e)}_{3,k,\sigma}(\tau,\tau') =
 \nonumber \\ &\quad
  \bar V_{k\bar\sigma}(\tau)\left[G_\sigma^{(e)}(\tau,\tau')
                                 -G_\sigma^{(2e)}(\tau,\tau')\right]
 \\
 &\quad -\sum_{k'}\left[
  \bar V_{k'\bar\sigma}(\tau)\Gamma^{(3e)}_{5,k'k,\sigma}(\tau,\tau') 
 -\bar V_{k'\sigma}^{\dagger}(\tau)\Gamma^{(3e)}_{6,k'k,\sigma}(\tau,\tau')
  \right]
 \nonumber
\end{align}
On the right-hand-side of these equations we now have new,
higher order GFs, $\Gamma^{(3e)}_{j,k'k,\sigma}$ defined by
the middle terms of Eqs.(\ref{defGam3e1})-(\ref{defGam3e6}).
GFs $\Gamma^{(2e)}$ and $\Gamma^{(3e)}$ take account of spin correlations
in the leads. Closure of the (in principle infinite) EOM chain is achieved 
assuming that higher-order spin correlations in the leads can be neglected.
Thus, following Ref.~\onlinecite{MeirWingreenLee_CB}, the terms $\Gamma^{(3e)}$ 
are expressed in terms of lower order GFs
\begin{align}
 \label{defGam3e1}
 \Gamma^{(3e)}_{1,k'k,\sigma}(\tau,\tau') &=
 -i <T_c \hat c_{k'\bar\sigma}^\dagger(\tau)\hat c_{k\sigma}(\tau)
    \hat d_{\bar\sigma}(\tau)\,\hat d_\sigma^\dagger(\tau') >
 \approx 0
 \\
 \label{defGam3e2}
 \Gamma^{(3e)}_{2,k'k,\sigma}(\tau,\tau') &=
 -i <T_c \hat c_{k\sigma}(\tau)\hat c_{k'\bar\sigma}(\tau)
    \hat d_{\bar\sigma}^\dagger(\tau)\,\hat d_\sigma^\dagger(\tau') >
 \approx 0
 \\
 \label{defGam3e3}
 \Gamma^{(3e)}_{3,k'k,\sigma}(\tau,\tau') &=
 -i <T_c \hat c_{k'\sigma}(\tau)\hat c_{k\bar\sigma}^\dagger(\tau)
    \hat d_{\bar\sigma}(\tau)\,\hat d_\sigma^\dagger(\tau') >
 \approx 0
 \\
 \label{defGam3e4}
 \Gamma^{(3e)}_{4,k'k,\sigma}(\tau,\tau') &=
 -i <T_c \hat c_{k\bar\sigma}^\dagger(\tau)\hat c_{k'\bar\sigma}(\tau)
    \hat d_\sigma(\tau)\,\hat d_\sigma^\dagger(\tau') >
 \\
 &\approx \delta_{k,k'}<\hat n_{k\bar\sigma}> G_\sigma^{(e)}(\tau,\tau')
 \nonumber \\
 \label{defGam3e5}
 \Gamma^{(3e)}_{5,k'k,\sigma}(\tau,\tau') &=
 -i <T_c \hat c_{k\bar\sigma}(\tau)\hat c^\dagger_{k'\bar\sigma}(\tau)
    \hat d_\sigma(\tau)\,\hat d_\sigma^\dagger(\tau') >
 \\
 &\approx \delta_{k,k'}\left[1-<\hat n_{k\bar\sigma}>\right]
         G_\sigma^{(e)}(\tau,\tau')
 \nonumber \\
 \label{defGam3e6}
 \Gamma^{(3e)}_{6,k'k,\sigma}(\tau,\tau') &=
 -i <T_c \hat c_{k\bar\sigma}(\tau)\hat d^\dagger_{\bar\sigma}(\tau)
    \hat c_{k'\sigma}(\tau)\,\hat d_\sigma^\dagger(\tau') >
 \approx 0
\end{align}
Now using (\ref{defGam3e1})-(\ref{defGam3e6}) in 
(\ref{eomGam2e1})-(\ref{eomGam2e3}) one can solve for 
$\Gamma^{(2e)}_{i,k,\sigma}$ ($i=\{1,2,3\}$) in terms of $G_\sigma^{(e)}$ and 
$G_\sigma^{(2e)}$
\begin{align}
 \Gamma^{(2e)}_{1,k,\sigma} &= g_{k,\sigma}\bar V_{k\sigma}\circ G_\sigma^{(2e)}
 \\
 \Gamma^{(2e)}_{2,k,\sigma} &= g_{k,\bar\sigma}^{(1)}
 \bar V_{k\bar\sigma}^\dagger \circ \left[ G_\sigma^{(2e)}
 - <\hat n_{k\bar\sigma}>G_\sigma^{(e)}\right]
 \\
 \Gamma^{(2e)}_{3,k,\sigma} &= g_{k,\bar\sigma}^{(2)}
 \bar V_{k\bar\sigma} \circ \left[ <\hat n_{k\bar\sigma}>G_\sigma^{(e)}
 - G_\sigma^{(2e)} \right]
\end{align}
where we have used short notation style with `$\circ$' implying 
convolution of two functions on the contour 
$(A\circ B)(\tau,\tau')=\int_c d\tau''\, A(\tau,\tau'')B(\tau'',\tau')$.
These solutions are substituted into (\ref{eomG2e})
which gives $G_\sigma^{(2e)}$ in terms of $G_\sigma^{(e)}$. Finally,
the last result together with (\ref{eomGam1e}) can be used in (\ref{eomGe}) 
to get equation for $G_\sigma^{(e)}$ in the form
\begin{equation}
 \label{Ge_MWL}
 G_\sigma^{(e)} = G_{2,\sigma}^{(e)}
 + U<\hat n_{\bar\sigma}> G_{2,\sigma}^{(e)}\circ G_{1,\sigma}^{(e)}
\end{equation}
$G_{i,\sigma}^{(e)}$ ($i=\{1,2,3,4\}$) are defined in 
Eqs.~(\ref{Gei})-(\ref{Ge4_m1}), while `self-energies' entering these 
definitions are given by Eqs.~(\ref{SEs0})-(\ref{SEs3}).

In order to simplify the structure we rewrite it in the form
\begin{equation}
 \label{Ge_revisited}
 G_\sigma^{(e)} = \left[1-<\hat n_{\bar\sigma}>\right] G_{2,\sigma}^{(e)}
 + <\hat n_{\bar\sigma}>\left\{G_{2,\sigma}^{(e)}
 + U G_{2,\sigma}^{(e)}\circ G_{1,\sigma}^{(e)}\right\}
\end{equation}
and note that
\begin{equation}
 \label{Ge_makesimple}
 \left\{\ldots\right\} = 
 G_{2,\sigma}^{(e)}\circ G_{1,\sigma}^{(e)}
 \left[\hat G^{-1}_{1,\sigma}+U\right] =
 G_{2,\sigma}^{(e)}\circ G_{1,\sigma}^{(e)} \hat G^{-1}_{4,\sigma} =
 G_{3,\sigma}^{(e)}
\end{equation}
The last equation follows from 
$\hat G^{-1}_{1,\sigma}\hat G^{-1}_{2,\sigma}=
 \hat G^{-1}_{4,\sigma}\hat G^{-1}_{3,\sigma}$.
Substitution of (\ref{Ge_makesimple}) into (\ref{Ge_revisited})
leads to (\ref{Ge}).
The retarded projection of (\ref{Ge}) is the final result of 
Ref.~\onlinecite{MeirWingreenLee_CB}.

\section{\label{B}Analytical expression for self-energy $\Sigma_{\sigma 1}^{(\infty)}$}
Here we derive analytical expressions for retarded and lesser projections
of $\Sigma_{\sigma 1}^{(\infty)}$, Eq.~(\ref{SEs1infty}), under Lorentzian
assumption for coupling between molecule and contacts, Eq.~(\ref{SigmaK_E}).
In the case of a dense continuum of states in the contacts (assumed here)
the sum in (\ref{SEs1infty}) can be converted to an integral, then retarded
and lesser projection of the SE (in energy domain) are
\begin{align}
 \label{SEs1rE}
 &\Sigma_{\sigma 1}^{(\infty)r}(E) =
 \sum_{K=L,R} \int_{-\infty}^{+\infty}\frac{d\epsilon}{2\pi}\,
 \frac{\Gamma_{K,\bar\sigma}(\epsilon)f_K(\epsilon)}
 {E-\epsilon-\varepsilon_\sigma+\varepsilon_{\bar\sigma}+i\gamma_{\bar\sigma}/2}
 \\
 \label{SEs1ltE}
 &\Sigma_{\sigma 1}^{(\infty)<}(E)
 \nonumber \\ &\quad =
 i \sum_{K=L,R} \int_{-\infty}^{+\infty}\frac{d\epsilon}{2\pi}\,
 \frac{\Gamma_{K,\bar\sigma}(\epsilon)f_K^2(\epsilon)}
 {\left(E-\epsilon-\varepsilon_\sigma+\varepsilon_{\bar\sigma}\right)^2
 +\left(\gamma_{\bar\sigma}/2\right)^2}
 \nonumber \\
 &\quad\approx 
 i \sum_{K=L,R} \int_{-\infty}^{+\infty}\frac{d\epsilon}{2\pi}\,
  \frac{\Gamma_{K,\bar\sigma}(\epsilon)f_K(\epsilon)}
 {\left(E-\epsilon-\varepsilon_\sigma+\varepsilon_{\bar\sigma}\right)^2
 +\left(\gamma_{\bar\sigma}/2\right)^2}
\end{align}
where second line of (\ref{SEs1ltE}) is correct for the case of 
$T\to 0$ (relevant for observation of the Kondo peak).

Introducing
\begin{equation}
 x = \beta(\epsilon-\mu_K)
\end{equation}
we arrive at integrals of the form
\begin{align}
 \label{intr}
 &\int_{-\infty}^{+\infty}dx\,\frac{1}{(x-x_1)(x-x_2)(x-x_3)}\frac{1}{e^x+1}
 \\
 \label{intlt}
 &\int_{-\infty}^{+\infty}dx\,\frac{1}{(x-x_1)(x-x_2)(x-x_3)(x-x_4)}
  \frac{1}{e^x+1}
\end{align}
for (\ref{SEs1rE}) and (\ref{SEs1ltE}) respectively, where
\begin{align}
 \label{x1}
 x_1 &= \beta\left(E_{K,\bar\sigma}^{(0)}-\mu_K+iW_{K,\bar\sigma}^{(0)}\right)
 \\
 \label{x2}
 x_2 &= x_1^{*}
 \\
 \label{x3}
 x_3 &= \beta\left(E-\varepsilon_\sigma+\varepsilon_{\bar\sigma}-\mu_K
        +i\frac{\gamma_{\bar\sigma}}{2}\right)
 \\
 \label{x4}
 x_4 &= x_3^{*}
\end{align}
with $K=L,R$. These integrals can be taken analytically by complex contour 
integration, the poles are at $x_1$, $x_2$, $x_3$ ($x_4$ in the case of
integral (\ref{intlt})), and also at $y_n\equiv i\pi(2n+1)$; 
$n=0,\pm 1,\pm 2,\ldots$
Performing the integration one arrives at the following expressions for
the SE projections
\begin{widetext}
\begin{align}
 \label{SEs1r_infty}
 &\Sigma_{\sigma 1}^{(\infty)r}(E) = 
 \sum_{K=L,R} \left\{ 
 \frac{i\Gamma_{K,\bar\sigma}^{(0)}W_{K,\bar\sigma}^{(0)}}{4\pi}
 \left[
 \frac{\psi^{*}(\frac{\pi-ix_1}{2\pi})}
      {E_2+i(W_{K,\bar\sigma}^{(0)}+\gamma_{\bar\sigma}/2)} -
 \frac{\psi(\frac{\pi-ix_1}{2\pi})}
      {E_2-i(W_{K,\bar\sigma}^{(0)}-\gamma_{\bar\sigma}/2)}
 \right] \right. 
 \\ &- 
 \frac{\Gamma_{K,\bar\sigma}^{(0)}\left[W_{K,\bar\sigma}^{(0)}\right]^2}{2\pi}
 \frac{\psi(\frac{\pi-ix_3}{2\pi})}
      {\left[E_2-i(W_{k,\bar\sigma}^{(0)}-\gamma_{\bar\sigma}/2)\right]
       \left[E_2+i(W_{k,\bar\sigma}^{(0)}+\gamma_{\bar\sigma}/2)\right]}
 + \left.
 \frac{\Gamma_{k,\bar\sigma}^{(0)}W_{k,\bar\sigma}^{(0)}}{4}
 \frac{1}{E_2+i(W_{k,\bar\sigma}^{(0)}+\gamma_{\bar\sigma}/2)}
 \right\}
 \nonumber
 \\
 \label{SEs1lt_infty}
 &\Sigma_{\sigma 1}^{(\infty)<}(E) = 
 i\sum_{K=L,R} \Gamma_{k,\bar\sigma}^{(0)}
 \left\{
 \frac{W_{k,\bar\sigma}^{(0)}\gamma_{\bar\sigma}}{2\pi}
 \mbox{Im}\left[
 \frac{\psi(\frac{\pi-ix_1}{2\pi})}
      {\left[E_2-i(W_{k,\bar\sigma}^{(0)}-\gamma_{\bar\sigma}/2)\right]
       \left[E_2-i(W_{k,\bar\sigma}^{(0)}+\gamma_{\bar\sigma}/2)\right]}
 \right] \right.
 \\
 & +
 \frac{\left[W_{k,\bar\sigma}^{(0)}\right]^2}{\pi}
 \mbox{Im}\left[
 \frac{\psi(\frac{\pi-ix_3}{2\pi})}
      {\left[E_2-i(W_{k,\bar\sigma}^{(0)}-\gamma_{\bar\sigma}/2)\right]
       \left[E_2+i(W_{k,\bar\sigma}^{(0)}+\gamma_{\bar\sigma}/2)\right]}
 \right]
 + \left.
 \frac{W_{k,\bar\sigma}^{(0)}}{2}
 \left(W_{k,\bar\sigma}^{(0)}+\frac{\gamma_{\bar\sigma}}{2}\right)
 \frac{1}{E_2^2+(W_{k,\bar\sigma}^{(0)}+\gamma_{\bar\sigma}/2)^2}
 \right\}
 \nonumber
\end{align}
with $E_2=E-\varepsilon_\sigma+\varepsilon_{\bar\sigma}-E_{k,\bar\sigma}^{(0)}$
and where $\psi$ is a Psi (digamma) function.\cite{AS}
Note that it is the second term in Eq.~(\ref{SEs1r_infty}) which is responsible
for Kondo effect appearance.
\end{widetext}


\begin{thebibliography}{99}
\bibitem{X_footnote}Note that the diagrammatic technique
becomes unusable also due to presence of shift operators $\hat X$ in
the transfer matrix elements. Their many-body character makes Wick's 
theorem inapplicable.
\bibitem{even_Kondo}Note however recent publications,\cite{Kiselev,Rosch}
   where possibility of observing a Kondo effect 
   at finite bias for the spin-singlet ground state was reported.
\bibitem{zero_order}In a zero order calculation no iterations are employed to
   attain convergence of electron-phonon coupling effect. This amounts to
   assuming that the phonon remains at its original thermal equilibrium
   even when coupled to the non-equilibrium electronic system.
\end{thebibliography}
\end{document}